\documentclass[fleqn,usenatbib]{mnras}

\usepackage{newtxtext,newtxmath}

\usepackage[T1]{fontenc}

\DeclareRobustCommand{\VAN}[3]{#2}
\let\VANthebibliography\thebibliography
\def\thebibliography{\DeclareRobustCommand{\VAN}[3]{##3}\VANthebibliography}

\usepackage{graphicx}	
\usepackage{amsmath}	

\title[Evolutionary pathways of galaxies and their SMBH]{Testing the evolutionary pathways of galaxies and their supermassive black holes and the impact of feedback from Active Galactic Nuclei via large multiwavelength datasets
}

\author[G. Mountrichas \& F. Shankar]{
George Mountrichas  $^{1}$\thanks{E-mail: gmountrichas@gmail.com} \&
Francesco Shankar$^{2}$\thanks{E-mail: F.Shankar@soton.ac.uk}
\\
$^{1}$Instituto de Fisica de Cantabria (CSIC-Universidad de Cantabria), Avenida de los Castros, 39005 Santander, Spain\\
$^{2}$School of Physics and Astronomy, University of Southampton, Highfield, SO17 1BJ, UK\\
}

\date{Accepted XXX. Received YYY; in original form ZZZ}

\pubyear{2015}

\begin{document}
\label{firstpage}
\pagerange{\pageref{firstpage}--\pageref{lastpage}}
\maketitle

\begin{abstract}
It is still a matter of intense debate how supermassive black holes (SMBH) grow, and the role played by feedback from active galactic nuclei (AGN) in the co-evolution of SMBHs and galaxies. To test the coevolution proposed by theoretical models, we compile a large AGN sample of 5639 X-ray detected AGN, over a wide redshift range, spanning nearly three orders of magnitude in X-ray luminosity. The AGN have been detected in the {\it{COSMOS-Legacy}}, the Bo$\rm \ddot{o}$tes, the XMM-{\it{XXL}} and the eFEDS fields. Using the specific star formation rate estimates, we split the AGN host galaxies into star forming (SF), starburst (SB) and quiescent (Q). Our results show that the AGN accretion is increased in SB systems compared to SF and Q. Our analysis reveals a mild increase of L$_X$ with M$_*$. The L$_X$/SFR ratio has a weak dependence on M$_*$, and at fixed M$_*$ it is highest in Q systems. The latter trend is mostly driven by the significant drop in SFR in the Q state. The measured strong variations in SFR from the SB/SF to Q mirror those predicted in merger models with AGN feedback. However, the observed mild variations in L$_X$ are at variance with the same models. We also study the evolution of SFR for a galaxy control sample and found that it is very similar to that of X-ray AGN. This suggests that either AGN play a minor role in the star formation quenching, or the relative timescales of the two processes are different.
\end{abstract}

\begin{keywords}
galaxies: active; Galaxies, galaxies: evolution ; Galaxies, galaxies: star formation ; Galaxies, X-rays: galaxies ; Resolved and unresolved sources as a function of wavelength
\end{keywords}



\section{Introduction}



Correlations have been observed in the local \citep[e.g.,][]{Magorrian1998, Ferrarese2000} and sometimes in the distant Universe \citep[e.g.,][]{Suh2020}, between the mass of the central supermassive black hole (SMBH) and several properties of the host galaxies, or even of the host dark matter haloes \citep[e.g.][]{Shankar2020}. These correlations are often not well established, possibly affected by biases \citep[e.g.,][]{Bernardi2007,Shankar2016, Shankar2017, Shankar2019}, varying with galaxy morphology \citep[e.g.,][]{Suh2019} and/or with the degree of activity of the central SMBH \citep[e.g.,][]{Reines2015} and often presenting significant dispersions \citep[e.g.,][]{Graham2016}. Although some degree of correlation undoubtedly exists between SMBH and host galaxy, pinning down the physical causes driving their evolution remains elusive just relying on current observational constraints of their scaling relations. Indeed, many theoretical models show strong degeneracies when comparing with present-day data \citep[e.g.,][]{Habouzit2021, Habouzit2022}.

A step further in our understanding of SMBH growth within galaxies can be performed by directly analysing samples of Active Galactic Nuclei (AGN), when the SMBHs become active at different accretion rates at different epochs, and knowledge of their host galaxies are available. Cold gas in the galaxy can indeed fuel star formation rate (SFR) as well as accrete onto the central SMBH triggering the AGN, thus correlations between SFR and SMBH accretion rate, as recorded in the AGN luminosity, are expected.

Semi-analytic models and hydrodynamical simulations have studied the star formation and the growth of the black hole in galaxy mergers. According to most of these studies, after the starburst (SB) phase that is triggered by the merger, both the star formation and the black hole accretion rate (BHAR) are quenched. The magnitude of the suppression depends on the mass of the galaxy, with more massive systems to present a steeper decrease of both the accretion rate and the star formation \citep[e.g.,][]{DiMatteo2005, Hopkins2012}. In the case of black holes the decrease of their  accretion rates is due to the accretion energy that is released. The star formation is suppressed by outflows generated by strong winds, produced by the AGN \citep[e.g.,][]{Debuhr2012} that remove or heat the star forming gas. However, this AGN feedback can work both ways \citep[e.g.,][]{Zinn2013}. For instance, in the late gas-poor phase, AGN feedback may quench star formation. In the gas-rich phase, AGN outflows can overcompress cold gas and thus enhance star formation \citep[e.g.,][]{Zubovas2013}.

From the observational point of view, previous works found that the average SMBH growth rate follows similar trends with stellar mass, M$_*$, and redshift as the star formation rates (SFRs) of their host galaxies \citep[e.g.,][]{Mullaney2012, Yang2018}. However, a stronger correlation has been found between the BHAR and the SFR rather than M$_*$, at least for AGN hosted by bulge dominated (BD) systems \citep{Yang2019}. Study of the BHAR/SFR ratio revealed that it scales weakly with M$_*$ \citep[e.g.,][]{Rodighiero2015, Yang2018, Carraro2020}, at least for star forming (SF) and starburst (SB) galaxies \citep[but see][]{Mullaney2012}, whereas a flat relation is found for Q systems \citep[][]{Carraro2020}. A drop of the BHAR/SFR ratio is found during the SB phase of galaxies \citep[][]{Rodighiero2015}, at odds with the theoretical predictions. Although, the AGN activity is enhanced in compact, SF galaxies, the bulk of the black hole growth takes place in extended, Q systems \citep{Aird2022}. Moreover, the mean L$_X$-M$_*$ relation is independent of the AGN duty cycle, but depends strongly on the shape, normalization and scatter of the underlying M$_{\rm BH}$-M$_*$ scaling relation and on the characteristic Eddington ratio \citep{Carraro2022}.

In this work, we use X-ray detected AGN in four fields, namely the the Bo$\rm \ddot{o}$tes, {\it{COSMOS-Legacy}}, the XMM-{\it{XXL}} and the eFEDS fields. Our sample consists of more than 5500 X-ray sources, spanning a redshift range of $\rm 0.5<z<2.5$ and nearly three orders of magnitude in X-ray luminosity, while their stellar mass ranges from $10.0<\rm log\,[M_*(M_\odot)]<12.0$. The samples used in our analysis are described in Sect. \ref{sec_data}. In Sect. \ref{sec_analysis}, we present the method we apply (SED fitting) to measure the host galaxy properties (SFR, M$_*$) and the various selection criteria we use to select only AGN with robust host galaxy measurements that also satisfy the mass completeness limits of each survey. We also describe how we classify sources into SB, SF and Q systems. Our goal is to study the X-ray luminosity (Sect. \ref{sec_lx_vs_mstar}) and the L$_X$/SFR ratio (Sect. \ref{sec_lxsfr_vs_mstar}), as a function M$_*$ and compare our observational results with the predictions of theoretical models (Sect. \ref{sec_comp_models}). Our conclusions are presented in Sect. \ref{sec_summary}. 

Throughout this work, we assume a flat $\Lambda$CDM cosmology with $H_ 0=70.4$\,Km\,s$^{-1}$\,Mpc$^{-1}$ and $\Omega _ M=0.272$ \citep{WMAP7}.

\section{Data}
\label{sec_data}

In our analysis, we use X-ray AGN detected in four fields, namely the Bo$\rm \ddot{o}$tes, the COSMOS-Legacy, the eFEDS and the XMM-XXL-N fields. The datasets are described in detail in \cite{Mountrichas2021c, Mountrichas2022a, Mountrichas2022b} and \cite{Masoura2018} papers, respectively. 


Very good photometric coverage is essential to obtain robust measurements of host galaxy properties (SFR, M$_*$), via spectral energy distribution (SED) fitting (see Sect. \ref{sec_analysis}). Therefore, all sources used in our analysis meet strict photometric selection requirements. Specifically, all sources have measurements in the following photometric bands: $u$, $g$, $r$, $i$, $z$, J, H, K, W1 or IRAC1, W2 or IRAC2 and W4 or MIPS/24, where IRAC1, IRAC2, MIPS/24 are the 3.6, 4.5 and 24\,$\mu$m photometric bands of $\it{Spitzer}$ and W1, W2 and W4 are the WISE photometric bands at 3.4, 4.6 and 22\,$\mu$m. Approximately $35\%$ of the AGN also have available far-IR photometry ($\it{Herschel}$).  \cite{Mountrichas2021b, Mountrichas2021c, Mountrichas2022a} used data from the XMM-${\it{XXL}}$, Bo$\rm \ddot{o}$tes and COSMOS fields and utilizing the CIGALE \citep{Boquien2019, Yang2020, Yang2022} SED fitting code, showed that lack of far-IR photometry does not affect the SFR calculations derived by CIGALE. 

We also exclude from our analysis sources with bad SED fits and unreliable host galaxy measurements (see Sect. \ref{sec_cigale}). Furthermore, only AGN that satisfy the mass completeness limits of each survey are included in our measurements. These limits have been calculated following the method described in \cite{Pozzetti2010}. The process is described in detail in Sections 3 of \cite{Mountrichas2021c, Mountrichas2022a, Mountrichas2022b}. Below we give a brief description of each dataset.

\begin{table*}
\caption{Number of X-ray AGN that meet the photometric, quality and mass completeness criteria (see text for more details) in each redshift interval.}
\centering
\setlength{\tabcolsep}{0.5mm}
\begin{tabular}{ccccc}
 \hline
field &total &$\rm 0.5<z<1.0$ & $\rm 1.0<z<1.5$ & $\rm 1.5<z<2.5$   \\
  \hline
Bo$\rm \ddot{o}$tes & 1020   & 590  & 298 & 132  \\
COSMOS              & 1073   & 328  & 272 & 473  \\
eFEDS               & 2860   & 1145 & 722 & 993  \\
XMM-XXL-N           & 686    & 409  & 118 & 159  \\
  \hline
all fields          & 5639   & 2472  & 1410 & 1757  \\

  \hline
\label{table_data}
\end{tabular}
\end{table*}

\begin{table*}
\caption{Number of X-ray AGN in each of the stellar mass and redshift bins used in our analysis.}
\centering
\setlength{\tabcolsep}{1.0mm}
\begin{tabular}{ccccc}
 \hline
$\rm log\,[M_*(M_\odot)]$ &total &$\rm 0.5<z<1.0$ & $\rm 1.0<z<1.5$ & $\rm 1.5<z<2.5$   \\
  \hline
$10.0-10.5$ & 230   & 184  & 9 & 37  \\
$10.5-11.0$ & 1299   & 809  & 282 & 208  \\
$11.0-11.5$ & 2797   & 1165 & 757 & 875  \\
$11.5-12.0$ & 1105    & 272  & 310 & 523  \\
  \hline
\label{table_data2}
\end{tabular}
\end{table*}

\subsection{COSMOS}
\label{sec_cosmos}

The {\it{COSMOS-Legacy}} survey \citep{Civano2016} is a 4.6\,Ms {\it{Chandra}} program that covers 2.2\,deg$^2$ of the COSMOS field \citep{Scoville2007}. The central area has been observed with an exposure time of $\approx 160$\,ks while the remaining area has an exposure time of $\approx 80$\,ks. The limiting depths are $2.2 \times 10^{-16}$, $1.5 \times 10^{-15}$ , and $8.9 \times 10^{-16}\,\rm erg\,cm^{-2}\,s^{-1}$ in the soft (0.5-2\,keV), hard (2-10\,keV), and full (0.5-10\,keV) bands, respectively. The X-ray catalogue includes 4016 sources. \cite{Marchesi2016} matched the X-ray sources with optical and infrared counterparts using the likelihood ratio technique \citep{Sutherland_and_Saunders1992}. Of the sources, 97\%  have an optical and IR counterpart and a photometric redshift (photoz) and $\approx 54\%$  have spectroscopic redshift (specz). Hardness ratios ($\rm HR=\frac{H-S}{H+S}$, where H and S are the net counts of the sources in the hard and soft band, respectively) were estimated for all X-ray sources using the Bayesian estimation of hardness ratios method \citep[BEHR;][]{Park2006}. The intrinsic column density, N$\rm _H$ , for each source was then calculated using its redshift and assuming an X-ray spectral power law with slope $\Gamma=1.8$. This information is available in the catalogue presented in \cite{Marchesi2016}.

We only use sources within both the COSMOS and UltraVISTA \citep{McCracken2012}  regions. UltraVISTA covers 1.38\,deg$^2$ of the COSMOS field \citep[][]{Laigle2016} and has deep near-infrared (NIR) observations ($J, H, K_s$ photometric bands) that allow us to derive more accurate host galaxy properties through SED fitting (see below). There are 1718 X-ray sources that lie within the UltraVISTA area of COSMOS. Out of these, 1073 satisfy the photometric criteria mentioned above, have reliable SED fits (see section \ref{sec_cigale}) and meet the mass completeness requirements \citep[see Sect. 3.4 in][]{Mountrichas2022a}.

The X-ray catalogue was cross-matched with the COSMOS photometric dataset produced by the HELP collaboration \citep{Shirley2019, Shirley2021} to construct the SEDs of the X-ray sources and fit them to obtain measurements of the host galaxy properties. HELP includes homogeneous and calibrated multiwavelength data from 23 of the premier extragalactic survey fields imaged by the {\it{Herschel}} Space Observatory which form the {\it{Herschel}} Extragalactic Legacy Project (HELP).

\subsection{BO$\rm \ddot{O}$TES}
\label{sec_bootes}

We also use X-ray AGN observed by the {\it{Chandra}} X-ray Observatory within the 9.3\,deg$^2$ Bo$\rm \ddot{o}$tes field of the NOAO Deep Wide-Field Survey (NDWFS). The catalogue is compiled and fully described in \cite{Masini2020}. It consists of 6891 X-ray point sources with an exposure time of about 10\,ks per XMM pointing and a limiting flux of $4.7\times 10^{-16}$, $1.5\times 10^{-16}$ and $\rm 9\times 10^{-16}\,erg\,cm^{-2}\,s^{-1}$, in the $\rm 0.5-7\,keV$, $\rm 0.5-2\,keV$ and $\rm 2-7\,keV$ energy bands, respectively. 2346 ($\sim 33\%$) of the X-ray sources in this catalogue, have available spectroscopic redshifts (specz). For the remaining sources, we use hybrid photometric redshifts \citep[photoz;][]{Duncan2018a, Duncan2018b, Duncan2019} that are available in the \cite{Masini2020} catalogue. The X-ray absorption of each X-ray AGN is available and is parameterized with $\rm N_H$. $\rm N_H$ has been calculated from HR estimations, by applying the BEHR method. A fixed Galactic absorption of $\rm N_{H, Gal}=1.04\times 10^{20}\,cm^{-2}$ is assumed. \cite{Mountrichas2021c} cross-matched the X-ray catalogue with the Bo$\rm \ddot{o}$tes photometric catalogue produced by HELP to enrich the dataset with photometry from optical to far-IR . The final X-ray  Bo$\rm \ddot{o}$tes sample used in our analysis consists of 1020 AGN. 

\subsection{eFEDS}
\label{sec_efeds}

In our analysis, we include X-ray AGN observed in the eFEDS field. The catalogue is presented in \cite{Brunner2022}. eROSITA \citep[extended ROentgen Survey with an Imaging Telescope Array;][]{Predehl2021} is the primary instrument on the Spektrum-Roentgen-Gamma (SRG) orbital observatory \citep{Sunyaev2021}. SRG was built to provide a sensitive, wide-field-of-view X-ray telescope with improved capabilities compared to those of XMM-Newton and Chandra, the two most sensitive targeting X-ray telescopes in operation. The dataset includes 27910 X-ray sources detected in the $0.2-2.3$\,keV energy band with a flux limit of $\approx 7 \times 10^{-15}$\,erg\,cm$^{-2}\,\rm s{^{-1}}$ in the $0.5-2.0$\,keV energy range. Details of the source detection are given in \cite{Brunner2022} (see their Sect 3.3, Appendix A and Fig. 3). In brief, their method is based on selecting source candidates according to the statistics of fitting source images with a point spread function (PSF-)convolved model. First, a preliminary catalogue is created that contains all the potential source candidates. Then, a background map is generated using the preliminary catalogue. Next, the preliminary catalogue is used as input to the PSF-fitting procedure, that selects reliable sources from this catalogue. By comparing the best-fit source model with a zero-flux (pure background) model, the algorithm calculates a detection likelihood $L$ for each source, defined as $L=$\,-ln$P$, where $P$ is the probability of the source being caused by random background fluctuation. The final catalogue consists of sources detected above a detection likelihood of 6 in the most sensitive $0.2-2.3$\,keV band \citep[see also Fig. 5 in][]{Salvato2022}. Various criteria are applied to exclude problematic sources from our analysis \citep[for a detail description see Sect. 2.1 in][]{Mountrichas2022b}. Furthermore, we restrict our sources to those within the KiDS+VIKING area \citep{Kuijken2019, Hildebrandt2020}. Near-infrared (NIR) photometry outside of this region is shallow which significantly affects the accuracy and reliability of the photoz calculations \citep[Sect. 6.1 in][]{Salvato2022}. This area encompasses  10294 extragalactic X-ray sources. \cite{Liu2022} performed a systematic X-ray spectral fitting analysis on all the X-ray systems, providing fluxes and luminosities ---among other X-ray properties--- for the eFEDS sources. We use their posterior median, intrinsic (absorption-corrected) X-ray fluxes in the $2-10$\,keV energy band. 

\cite{Mountrichas2022b} restricted the X-ray catalogue to $\rm 0.5<z<1.5$. This was due to the unavailability of a large reference (non-AGN) sample at higher redshifts, with which they could compare the SFR of their AGN. In our analysis, we extend the X-ray catalogue to also include AGN at $\rm 1.5<z<2.5$, applying the same photometric and quality criteria as in \cite{Mountrichas2022b}. We also select only X-ray sources that satisfy the mass completeness limits of the field at this redshift range, following the method of \cite{Pozzetti2010} \citep[for more details see Sect. 3.4 in][]{Mountrichas2022b}. Our X-ray AGN catalogue consists of 2,860 sources within a redshift range of $\rm 0.5<z<2.5$.

\subsection{XMM-XXL}
\label{sec_xxl}

The XMM-Newton {\it{XXL}} survey \citep[XMM-{\it{XXL}};][]{Pierre2016} is a medium-depth X-ray survey that covers a total area of 50\,deg$^2$ split into two fields equal in size, the XMM-{\it{XXL}} North (XXL-N) and the XMM-{\it{XXL}} South (XXL-S). The {\it{XXL}}-N sample consists of 8445 X-ray sources. Of these X-ray sources, 5294 have SDSS counterparts and 2512 have reliable spectroscopy \citep{Menzel2016, Liu2016}. Mid-IR and near-IR was obtained following the likelihood ratio method \citep{Sutherland_and_Saunders1992} as implemented in \cite{Georgakakis_Nandra2011}. For more details on the reduction of the {\it{XMM}} observations and the IR identifications of the X-ray sources, see \cite{Georgakakis2017b}. The X-ray absorption of the AGN has been calculated in \cite{Masoura2021}, by calculating the HR of the sources using the BEHR method \citep[for details see Sect. 3.2 in][]{Masoura2021}. For those sources with no specz measurement, we use their photoz calculations presented in \cite{Masoura2018}. For their estimation, a machine-learning technique has been applied \citep[TPZ;][]{Kind2013}, as described in \cite{Mountrichas2017b, Ruiz2018}.

In our analysis, we use the X-ray sample presented in \cite{Masoura2018}. However, for consistency with the other X-ray datasets used in our analysis, we apply the same photometric criteria mentioned in Sect. \ref{sec_data} as well as the same quality criteria (see Sect. \ref{sec_cigale}). Furthermore, we calculate the mass completeness limits of the field at the three redshift intervals used in our study and use only AGN that have M$_*$ above these limits. For this purpose, we apply the method described in \cite{Pozzetti2010}, utilizing equation (1) in \cite{Mountrichas2021c} and using the $K$ band with $m_{lim}=20.0$\,mag \citep{Georgakakis2017b}. Our calculations show that the mass completeness of the sample is defined at $\rm log\,[M_{*,95\%lim}(M_\odot)]= 10.55, 11.00, 11.20$ at $\rm 0.5<z<1.0$, $\rm 1.0<z<1.5$ $\rm and \, 1.5<z<2.5$, respectively. Application of these criteria results in 686 X-ray AGN in {\it{XXL}}-N.

\begin{table*}
\caption{Number of X-ray AGN that live in quiescent (Q), star-forming (SF) and starburst (SB) galaxies, in the three redshift intervals used in our analysis. Only AGN with $10.0<\rm log\,[M_*(M_\odot)]<12.0$ are included. In the parentheses we quote the fraction of each AGN population in the corresponding redshift interval.}
\centering
\setlength{\tabcolsep}{1.0mm}
\begin{tabular}{ccccc}
 \hline
&total &$\rm 0.5<z<1.0$ & $\rm 1.0<z<1.5$ & $\rm 1.5<z<2.5$   \\
  \hline
SF & 3575   &  1775 (73\%) & 920 (68\%) & 880 (53.5\%)  \\
SB & 939   & 181 (7.5\%)  & 261 (19\%) & 497 (30.3\%) \\
Q & 917   & 474 (19.5\%)& 177 (13\%) & 266 (16.2\%)  \\
  \hline
\label{table_class}
\end{tabular}
\end{table*}

\subsection{Final sample}
\label{sec_final_sample}

\begin{figure}
\centering
  \includegraphics[width=0.8\columnwidth, height=6cm]{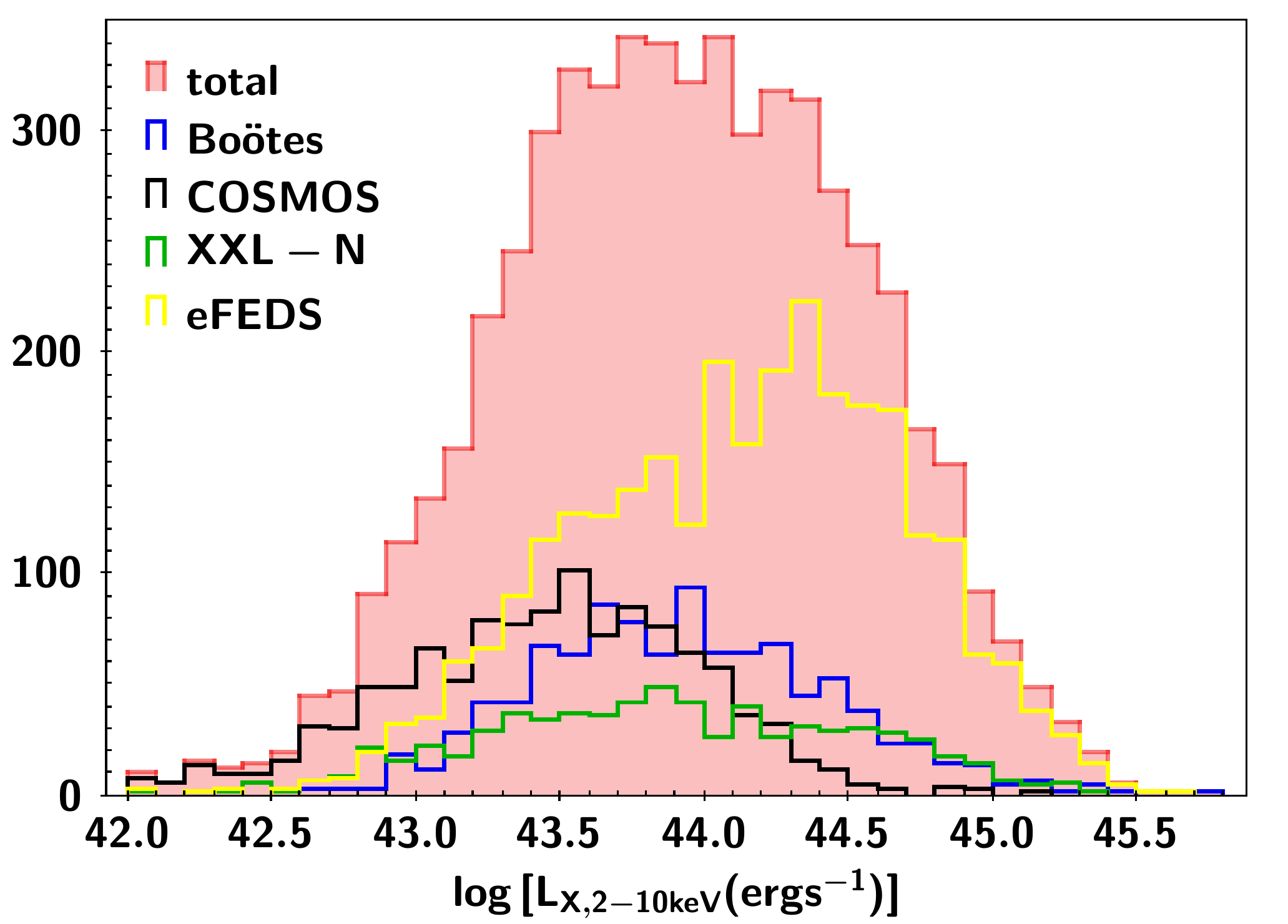} 
  \caption{The intrinsic L$_X$ distributions of the total sample used in our analysis (red shaded area) and of AGN in the four individual fields. eFEDS has the most luminous AGN while the COSMOS field consists, mainly, of low-to-moderate luminosity sources. }
  \label{fig_lx_distrib}
\end{figure}

\begin{figure}
\centering
  \includegraphics[width=0.8\columnwidth, height=6cm]{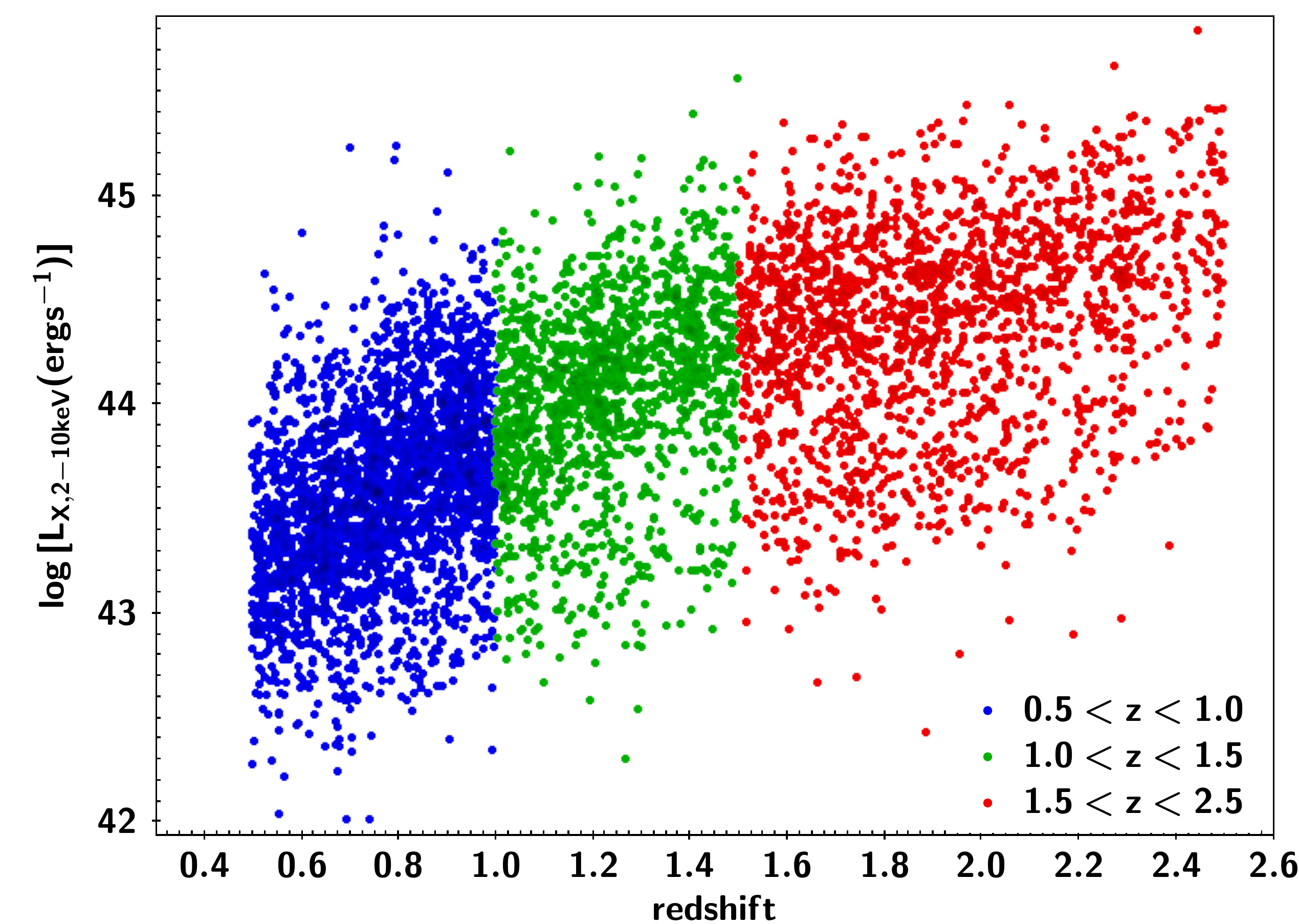} 
  \caption{The intrinsic L$_X$ as a function of redshift. The datapoints are colour-coded following the three redshift intervals used in our measurements.}
  \label{fig_lx_redz}
\end{figure}

Our final AGN dataset consists of 5639 X-ray sources, covering a redshift range of $\rm 0.5<z<2.5$. Table \ref{table_data}, presents the number of sources in each field and redshift interval. Fig. \ref{fig_lx_distrib} presents the intrinsic (absorption corrected) L$_X$ distribution of the AGN used in this study. Our sources span more than three orders of magnitude in L$_X$. The eFEDS field contributes the majority of the luminous ($\rm log\,[L_{X,2-10keV}(ergs^{-1})]>44$) sources used in our work, while most of the AGN in the COSMOS field are low-to-moderate luminosity. Fig. \ref{fig_lx_redz} presents the X-ray luminosity as a function of redshift, for the three redshift bins used in our analysis. In our analysis, we also split the X-ray sources into four M$_*$ bins. Table \ref{table_data2}, presents the number of sources in each M$_*$ and redshift bin.

\section{Analysis}
\label{sec_analysis}

In this section, we describe the SED fitting analysis applied to calculate the AGN host galaxy properties and how this information is used to classify the systems that AGN reside into quiescent, star-forming and starburst.


\subsection{SED fitting}
\label{sec_cigale}


The host galaxy properties of the AGN in the the Bo$\rm \ddot{o}$tes, {\it{COSMOS-Legacy}},  and the eFEDS fields, have been measured in \cite{Mountrichas2021c, Mountrichas2022a, Mountrichas2022b}, respectively, by applying SED fitting. For that purpose, the CIGALE algorithm \citep{Boquien2019, Yang2020, Yang2022}  has been utilized. CIGALE allows inclusion of the X-ray flux in the fitting process and has the ability to account for the extinction of the ultraviolet (UV) and optical emission in the poles of AGN \citep{Yang2020, Mountrichas2021a, Buat2021}. 

The same templates and parameter space has been used to fit the AGN in the different datasets (Sect. \ref{sec_data}). This minimises any systematic effects that may be introduced due to the different modules and parametric grid used in the fitting process. As already noted, in the case of the {\it{XXL}}-N sources, we do not use the SFR and M$_*$ measurements of \cite{Masoura2018}. Instead, we have made new runs of CIGALE for these AGN, using the same SED fitting analysis, as for the other three fields.

A detailed description of the fitting process is provided in \cite{Mountrichas2021c, Mountrichas2022a, Mountrichas2022b}. In brief, a delayed star formation history (SFH) model with a function form $\rm SFR\propto t \times exp(-t/\tau$) is used to fit the galaxy component. A star formation burst is included \citep{Ciesla2017, Malek2018, Buat2019} as a constant ongoing period of star formation of 50\,Myr. The \cite{Bruzual_charlot2003} single stellar population template is used to model the stellar emission. Stellar emission is attenuated following \cite{Charlot_Fall_2000}. The dust heated by stars is modelled following \cite{Dale2014}. The SKIRTOR template \citep{Stalevski2012, Stalevski2016} is used for the AGN emission. The values for the various parameters of the SED fitting are similar for the AGN in all the four fields used in this work and are presented in Tables 1 in \cite{Mountrichas2021c, Mountrichas2022a, Mountrichas2022b}.

To exclude sources that are badly fitted and have unreliable host galaxy measurements, we follow the criteria applied in previous studies \citep[e.g.,][]{Mountrichas2021b, Mountrichas2021c, Buat2021, Mountrichas2022a, Mountrichas2022b}. Specifically, we reject AGN with a reduced $\chi ^2 >5$. We also exclude sources for which CIGALE could not constrain their SFR and M$_*$. For that purpose, we use the two estimates that CIGALE provides for each galaxy property. One value is evaluated from the best-fit model and the other weighs all models allowed by the parametric grid, with the best-fit model having the heaviest weight \citep{Boquien2019}. A large difference between these two values indicates that the probability density function (PDF) is asymmetric and a simple model for the errors is not valid. Thus, in our analysis we only keep sources with  $\rm \frac{1}{5}\leq \frac{SFR_{best}}{SFR_{bayes}} \leq 5$ and $\rm \frac{1}{5}\leq \frac{M_{*, best}}{M_{*, bayes}} \leq 5$, where SFR$\rm _{best}$ and M$\rm _{*, best}$ are the best-fit values of SFR and M$_*$, respectively, and SFR$\rm _{bayes}$ and M$\rm _{*, bayes}$ are the Bayesian values estimated by CIGALE \citep[e.g.,][]{Mountrichas2021b, Mountrichas2021c, Buat2021, Mountrichas2022a, Mountrichas2022b}.

\begin{figure}
\centering
  \includegraphics[width=0.95\columnwidth, height=7cm]{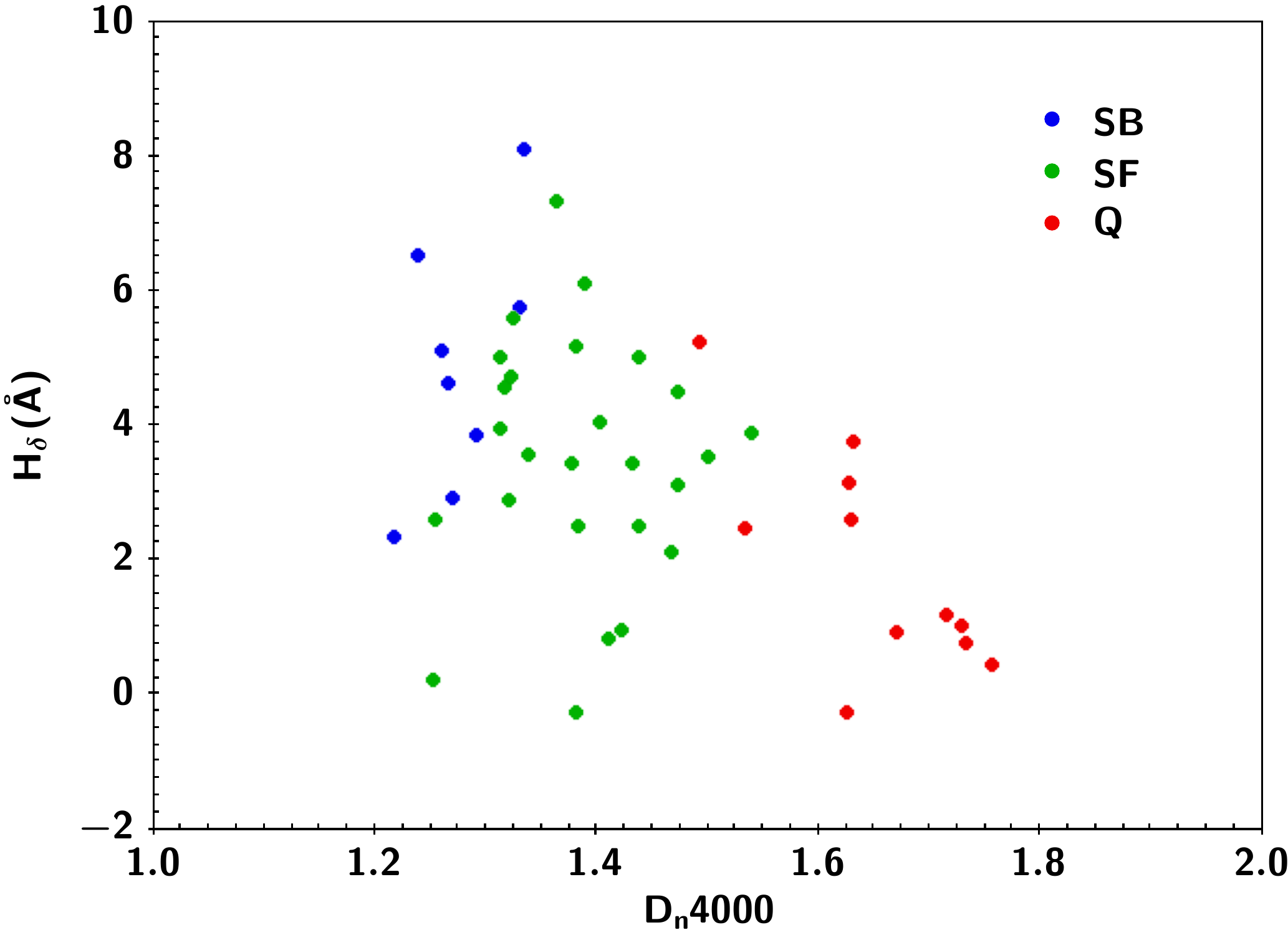} 
  \caption{H$_\delta$ vs. D$_n$4000 for the 52 of the X-ray AGN from our analysis that are also included in the LEGA-C catalogue. AGN hosted by SB galaxies have the youngest stellar populations (smallest D$_n$4000 values) and have experienced recent star formation bursts (high H$_\delta$ values). Q systems have the oldest stars (highest D$_n$4000 values) and present no signs of recent bursts (smallest H$_\delta$ values). AGN that live in SF galaxies have intermediate H$_\delta$ and D$_n$4000 values.}
  \label{fig_hdelta_dn4000}
\end{figure}

\subsection{Classification of sources}
\label{sec_classification}

To identify AGN host galaxies that are quiescent, we use the criterion presented in \cite{Mountrichas2021c, Mountrichas2022a, Mountrichas2022b}. This criterion uses the distribution of the specific SFR ($\rm sSFR=\frac{SFR}{M_*}$) of each dataset, at different redshift intervals. These distributions present a lower, second peak at low sSFR values. The galaxies that populate these lower peaks are classified as quiescent. Systems that have sSFR that is 0.6\,dex higher than the mean value of the dataset are considered starburst \citep[e.g.,][]{Rodighiero2011, Rodighiero2015}. The remaining of the AGN population is then classified as star-forming. To examine whether our results and conclusions, presented in the next Sections, are sensitive to the classification method applied, we have also used different criteria for the AGN host galaxy classification. For instance, we identify as quiescent sources that have sSFR values 0.3\,dex below the mean value and as SB systems that have 0.3\,dex above the mean sSFR value. We confirm that our results and conclusion are not affected by the classification method applied.  

Furthermore, we examine whether our classification is consistent with the ages of the stellar populations of the AGN hosts. For that purpose, we use the information included in the Large Early Galaxy Astrophysics Census (LEGA-C) catalogue. The LEGA-C catalogue includes data obtained from the LEGA-C survey \citep{Wel2021, Straatman2016}. The third data release contains galaxy spectra for 3741 unique sources that cover a redshift range from 0.6 to 1.3. The sources lie within the UltraVISTA region of the COSMOS field. The catalogue includes measurements for two stellar age sensitive tracers, the equivalent width (EW) of H$\delta$ absorption and the D$_n$4000 index \citep[][]{Worthey1997, Balogh1999}. D$_n$4000 is small for young stellar populations and large for old, metal rich galaxies. The EW of H$_\delta$ rises rapidly in the first few hundred Myrs after a burst of star formation, when O- and B-type stars dominate the spectrum and then decreases when A-type stars fade \citep[e.g.,][]{Kauffmann2003, Wu2018, Mountrichas2022c}. We cross-match our X-ray dataset with the LEGA-C catalogue, using a radius of 1\arcsec and the optical coordinates provided in each catalogue. There are 52 AGN that are included in both catalogues, after applying several criteria to exclude sources with unreliable measurements \citep[for more details see Sect. 2.2.3 in][]{Mountrichas2022c}.  Our goal is to examine how our classified AGN host galaxies are distributed in the H$_\delta$-D$_n$4000 space. This is shown in Fig. \ref{fig_hdelta_dn4000}. AGN hosted by SB galaxies have the youngest stellar populations (smallest D$_n$4000 values) and have experienced (recent) star formation bursts (high H$_\delta$ values). Q systems have the oldest stars (highest D$_n$4000 values) and present no signs of recent bursts (smallest H$_\delta$ values). AGN that live in SF galaxies have intermediate H$_\delta$ and D$_n$4000 values. Although this subset is small, the results of this exercise indicate that the criteria we have applied to classify AGN host galaxies into SB, SF and Q are robust. 

Table \ref{table_class} presents the number of Q, SB and SF galaxies that host X-ray AGN, in the three redshift intervals used in our analysis. Only sources with $10.0<\rm log\,[M_*(M_\odot)]<12.0$ are taken into consideration. As expected, the fraction of AGN hosted by SB galaxies increases as we move to higher redshifts while the fraction of quiescent systems increases at lower redshifts \citep[e.g.][]{Shimizu2015, Koutoulidis2022}. However, the bulk of the accretion density of the Universe is associated with star-forming systems, at all redshift ranges probed by our sample, in agreement with previous studies \citep[e.g.,][]{Georgakakis2014, Rodighiero2015}. At $\rm 0.5<z<1.5$ the fraction of AGN hosted by quiescent systems is similar to that found by \cite{Mountrichas2021b} ($\sim 12\%$, see their Fig. 6). At higher redshifts ($\rm 1.5<z<2.5$) the fraction of quiescent galaxies that host AGN ($\sim 16\%$) is consistent with that found by \cite{Rodighiero2015} ($\sim 11\%$). Similar fractions of quiescent galaxies ($15-20\%$) has also been found in non-AGN systems \citep[e.g.,][]{Fontana2009}. Regarding, AGN hosted by SB galaxies, in our dataset this fraction is significantly higher compared to that quoted by \cite{Rodighiero2015} ($\sim 30\%$ vs. $\sim 6\%$). Albeit, our sample spans significantly higher X-ray luminosities compared to theirs, in this redshift regime ($\rm 43.5<log\,[L_{X,2-10keV}(ergs^{-1})]<45$ vs. $\rm 41.8<log\,[L_{X,2-10keV}(ergs^{-1})]<43.6$) and AGN that span such high X-ray luminosities have been associated with increased SFR compared to less luminous sources \citep[e.g.,][]{Masoura2021, Mountrichas2021c, Mountrichas2022b, Pouliasis2022}.

\begin{figure}
\centering
  \includegraphics[width=0.8\columnwidth, height=6cm]{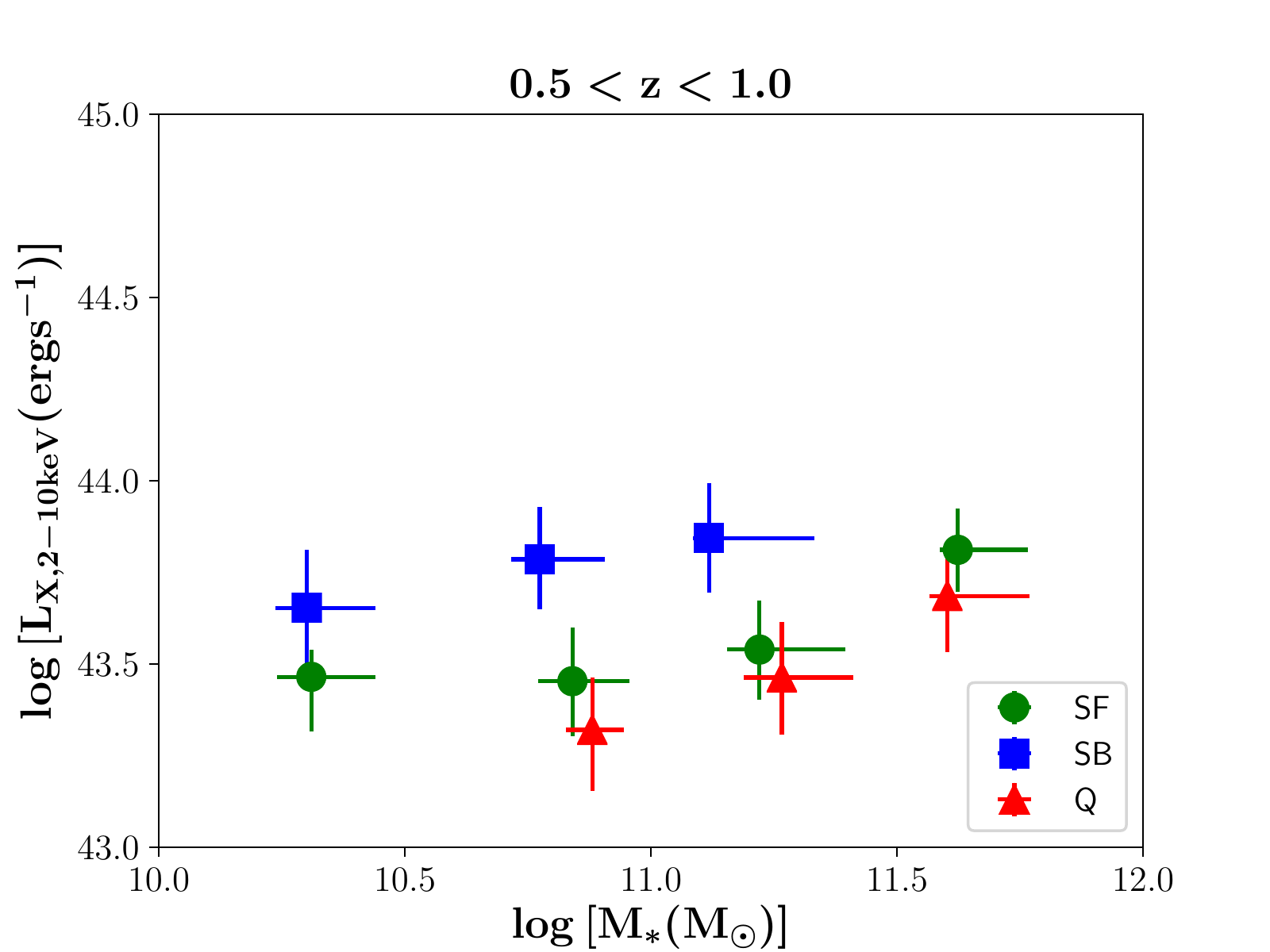} 
  \includegraphics[width=0.8\columnwidth, height=6cm]{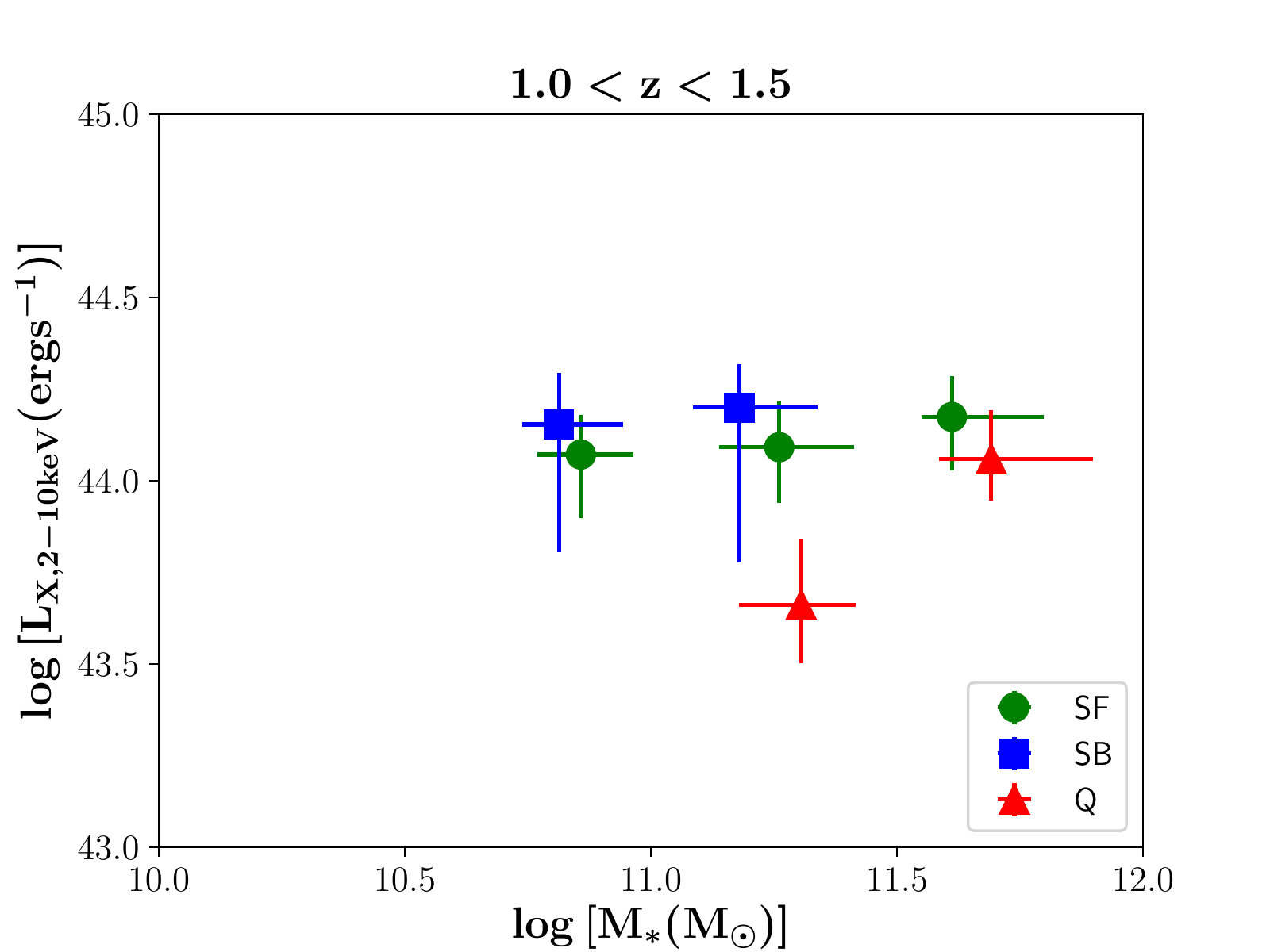} 
  \includegraphics[width=0.8\columnwidth, height=6cm]{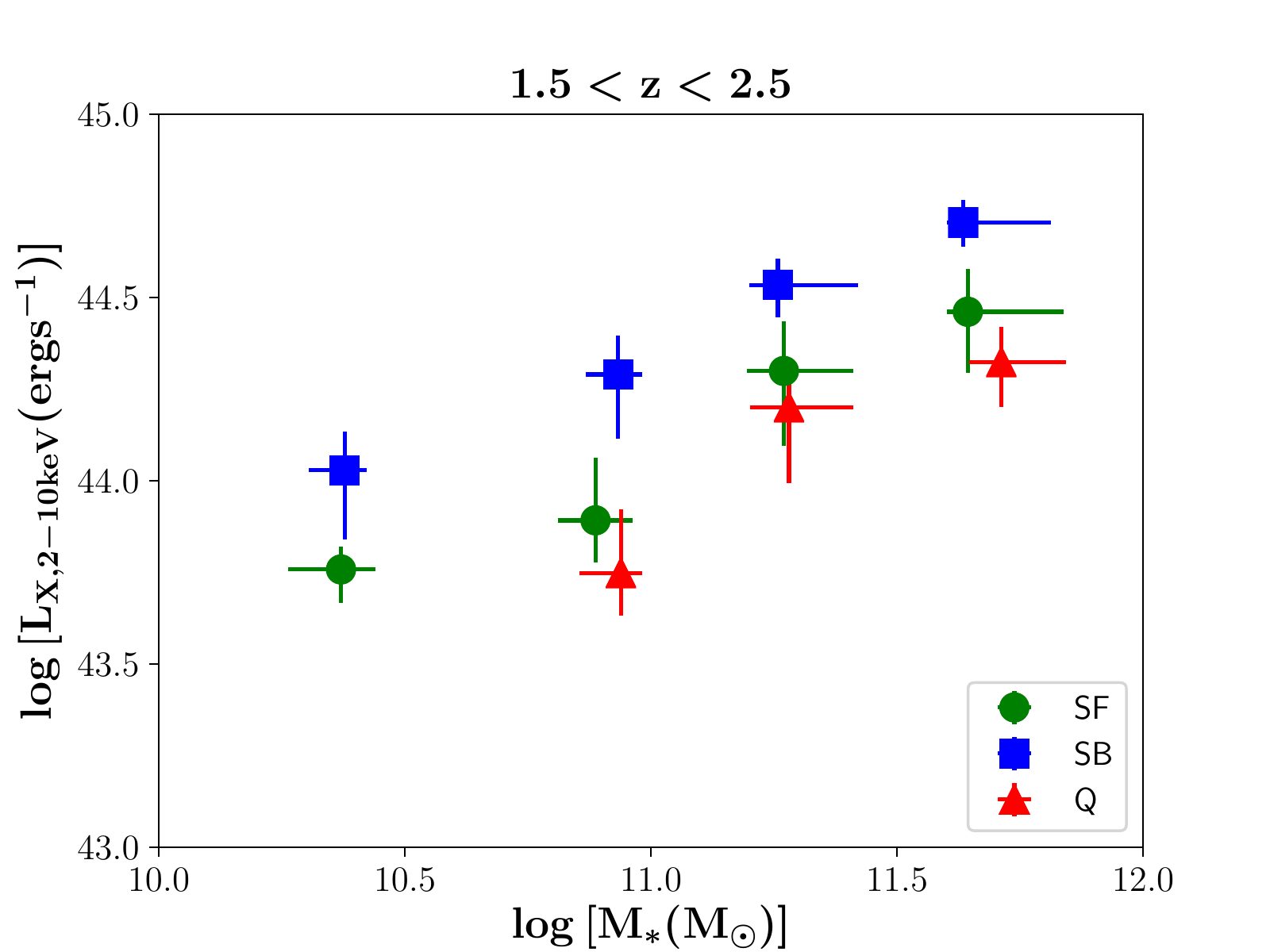}  
  \caption{L$_X$ as a function of M$_*$. The results are colour-coded based on the classification of the sources. Each panel corresponds to a different redshift interval. Errors have been calculated using bootstrap resampling. SB systems showcase an increased median L$_X$ compared to SF galaxies and Q systems show a deficit of L$_X$ compared to SF galaxies, at similar M$_*$. A mild increase of L$_X$ (by $\sim 0.5$\,dex) with M$_*$ is found for all AGN host galaxy classifications, at all redshifts spanned by our dataset.}
  \label{fig_lx_mstar_all}
\end{figure} 

\begin{figure}
\centering
  \includegraphics[width=0.8\columnwidth, height=6cm]{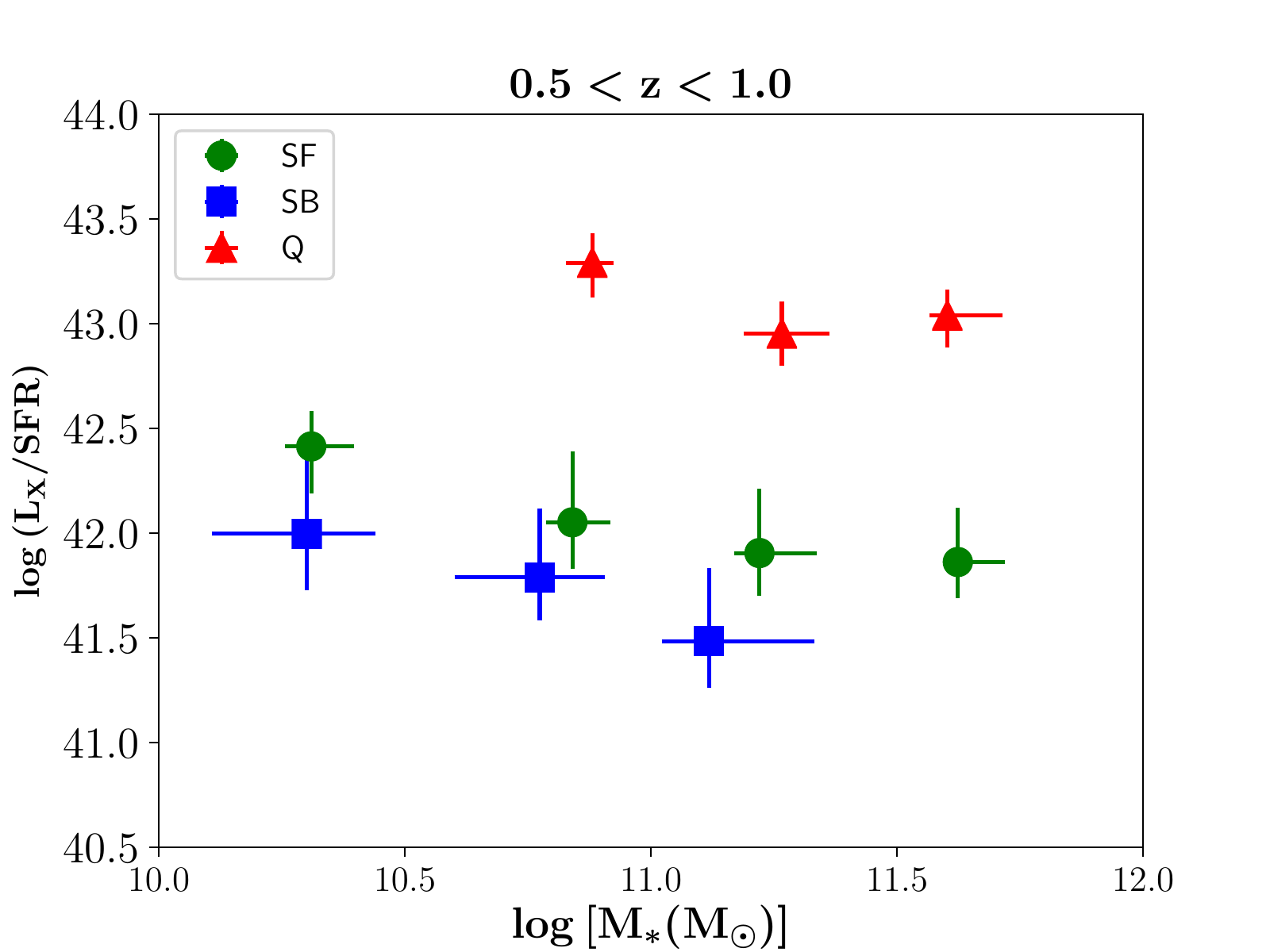} 
  \includegraphics[width=0.8\columnwidth, height=6cm]{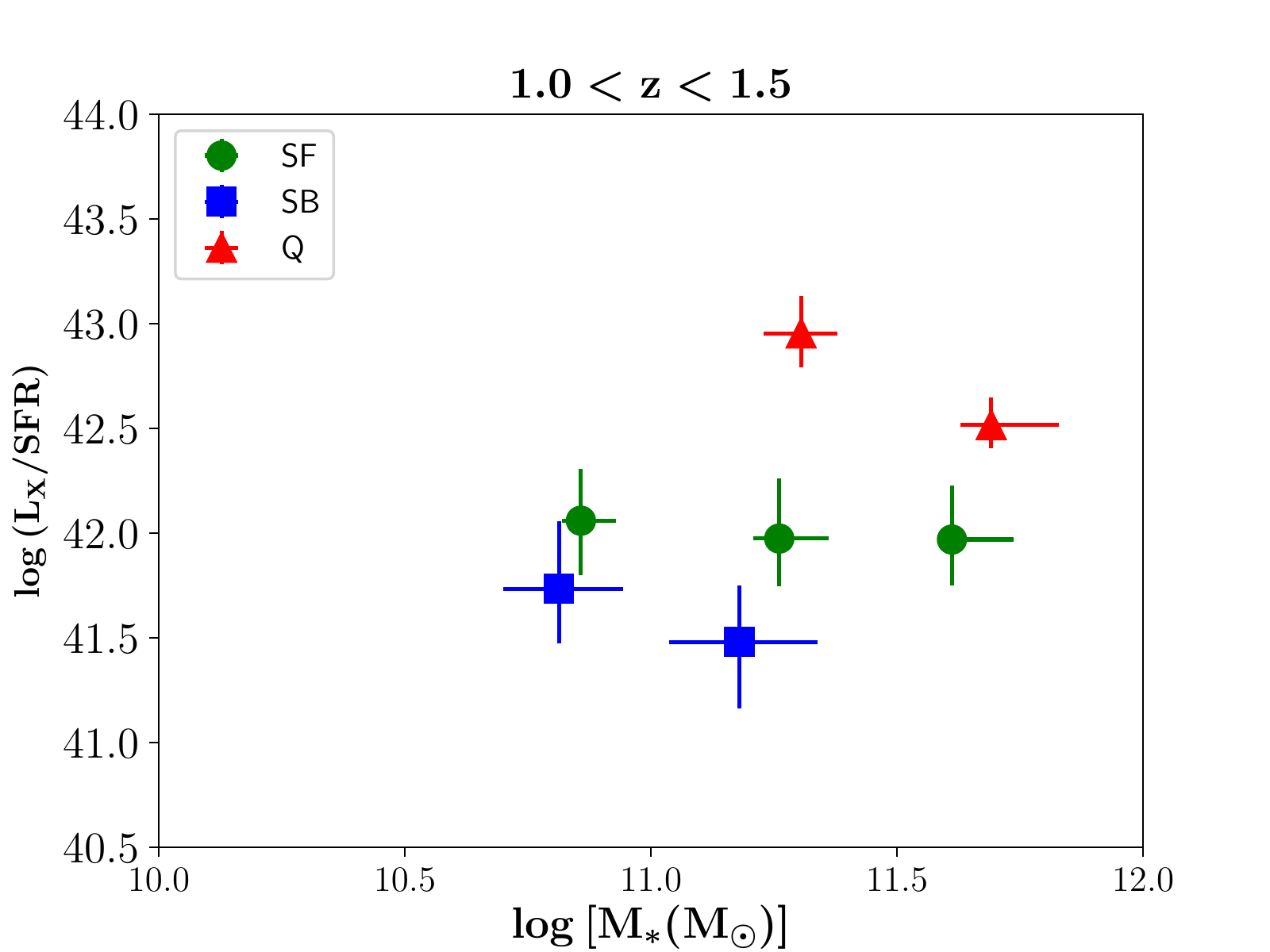} 
  \includegraphics[width=0.8\columnwidth, height=6cm]{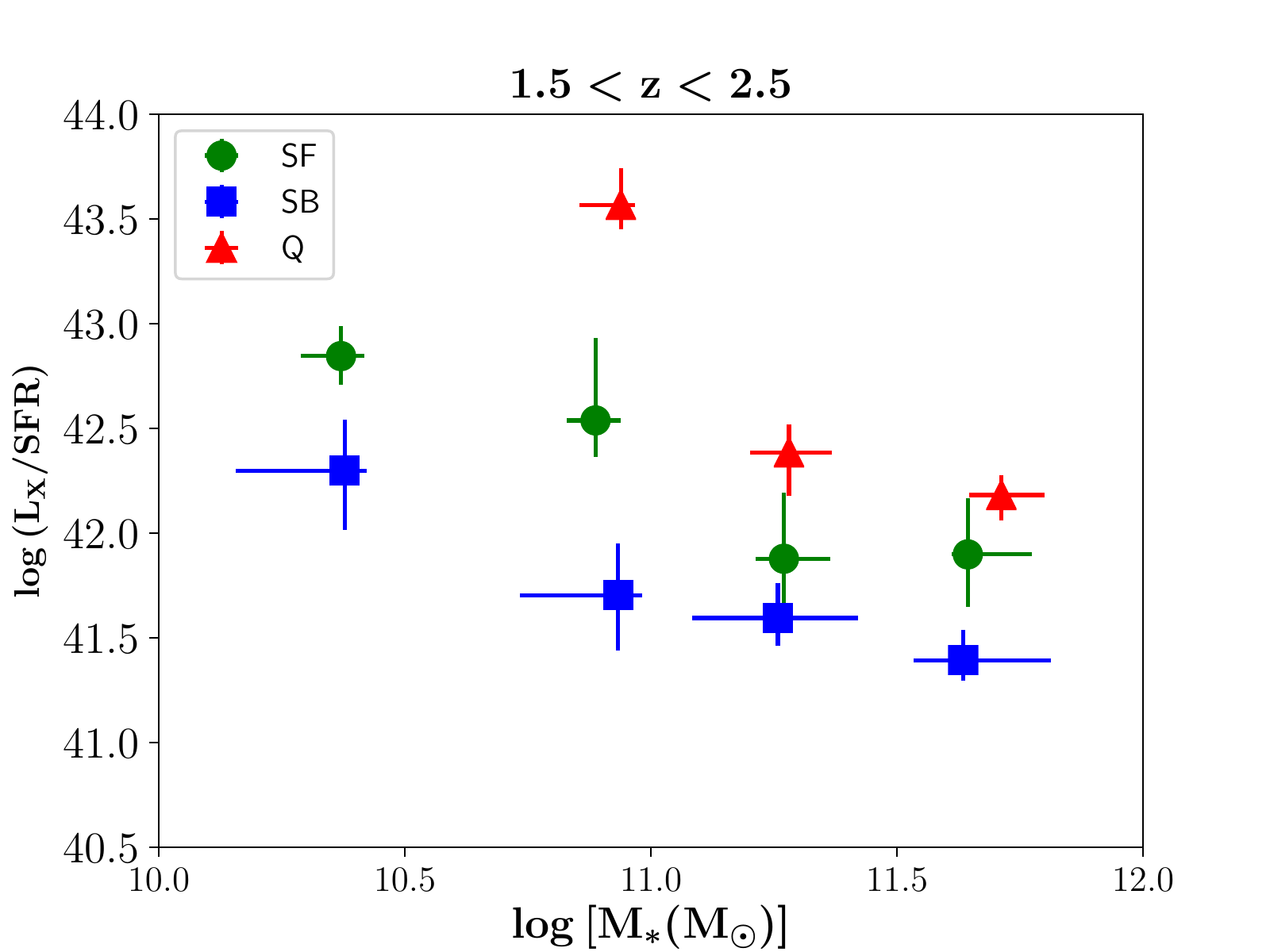}  
  \caption{L$_X$/SFR as a function of M$_*$. The amplitude of L$_X$/SFR is higher for Q compared to SF and SB and presents only a weak dependence on stellar mass.}
  \label{fig_lxsfr_mstar_all}
\end{figure} 

\begin{figure}
\centering
  \includegraphics[width=0.8\columnwidth, height=6cm]{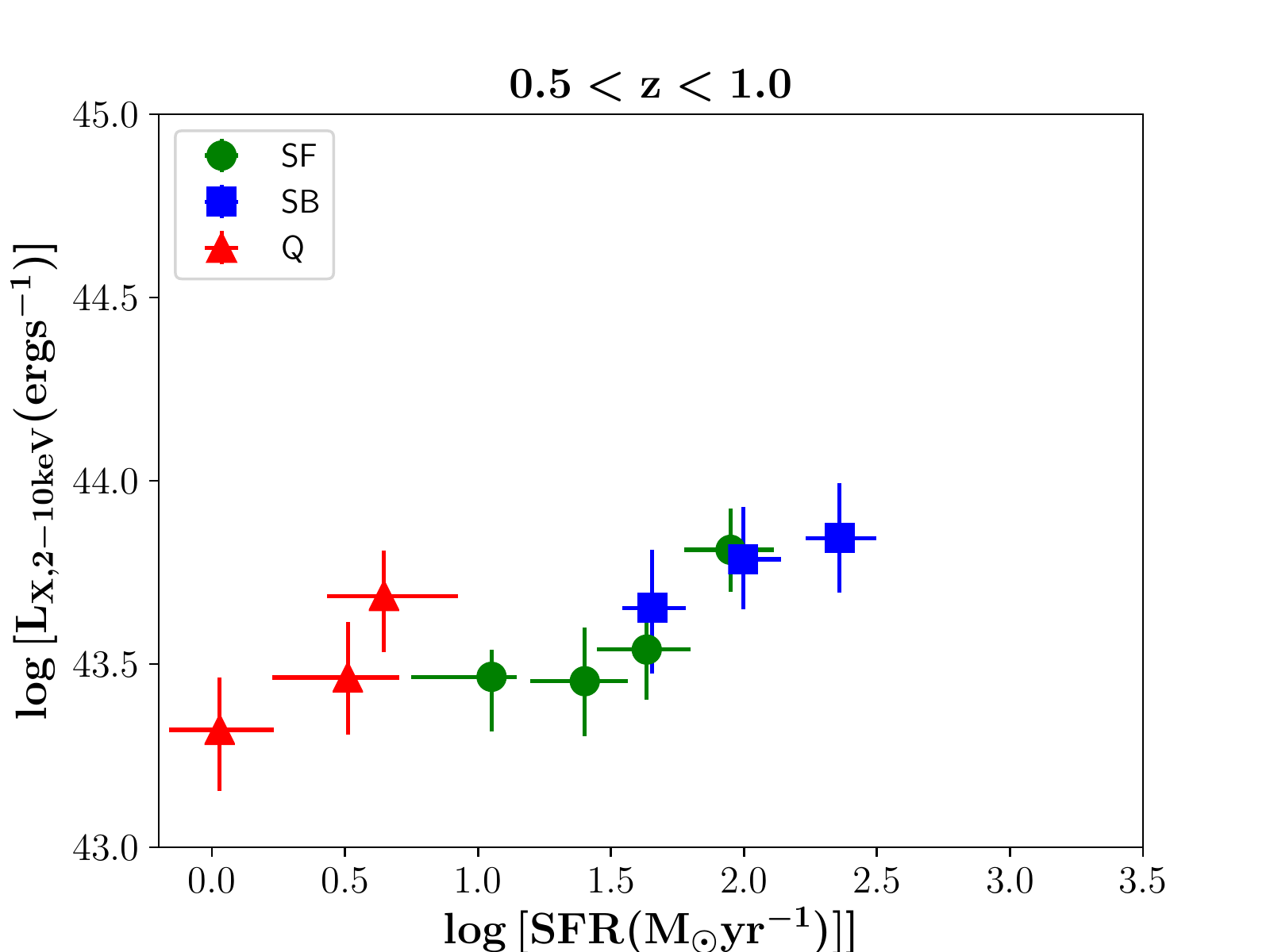} 
  \includegraphics[width=0.8\columnwidth, height=6cm]{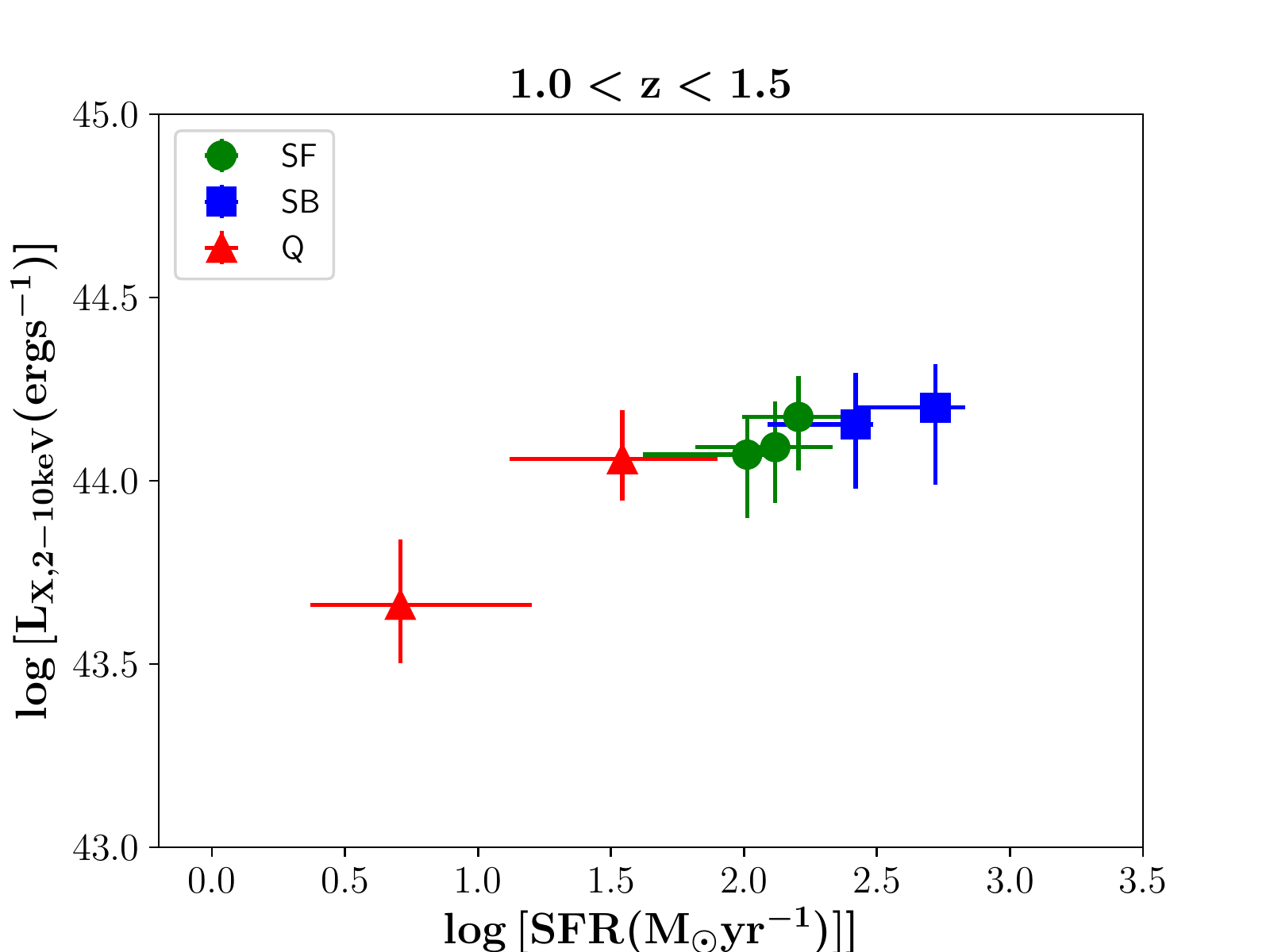} 
  \includegraphics[width=0.8\columnwidth, height=6cm]{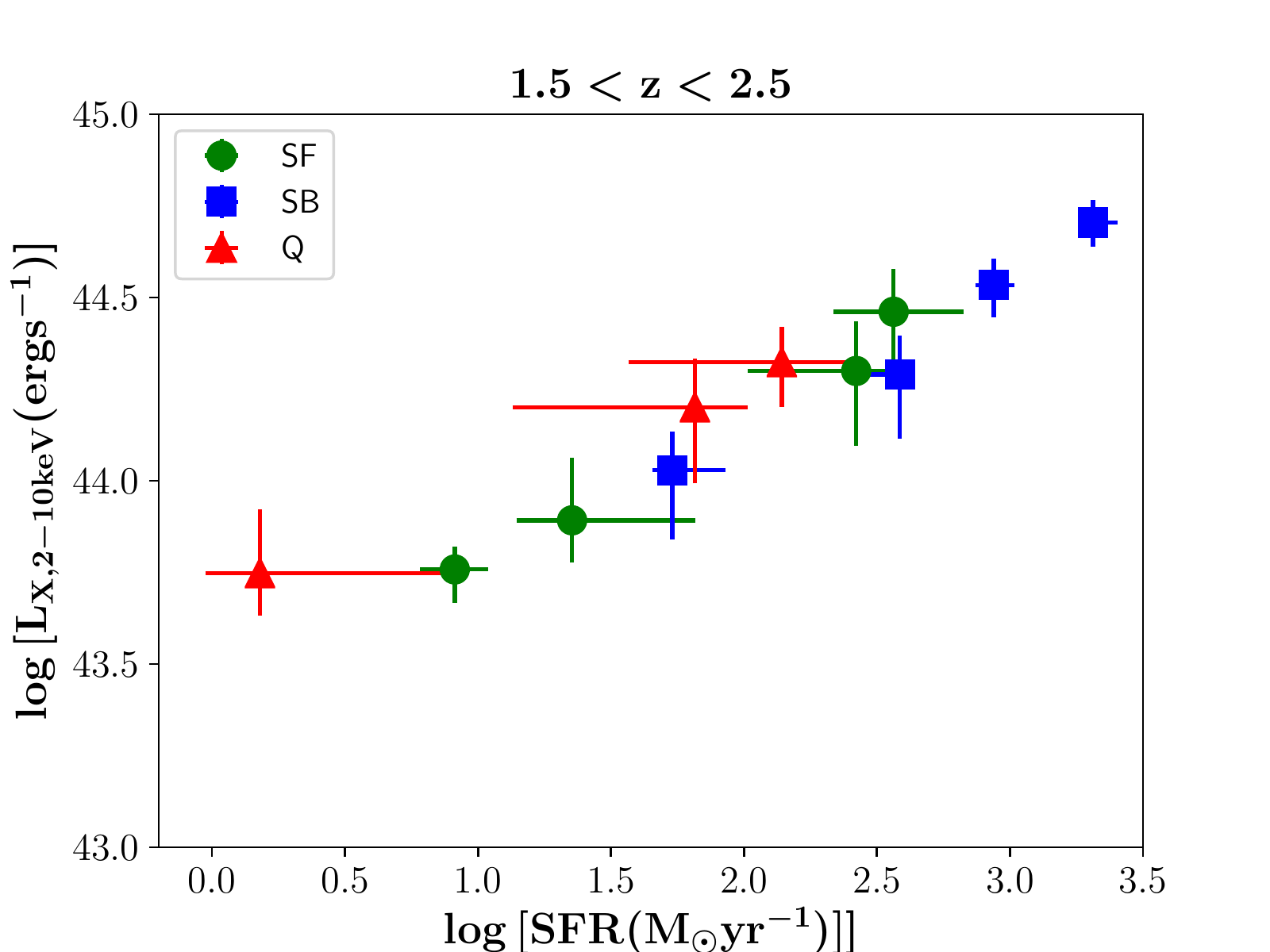}  
  \caption{L$_X$ as a function of SFR for SB, SF and Q AGN host galaxies, at three redshift intervals. At all redshift ranges, L$_X$ increases by a smaller factor compared to the increase of SFR.} 
  \label{fig_lx_sfr_check}
\end{figure}


\section{Results}

In this Section, we study the AGN accretion (L$_X$) and the L$_X$/SFR ratio as a function of stellar mass. All measurements presented are the median values of the properties in each bin. Bins that include less than 20 sources are excluded from our analysis. All errors presented have been calculated using a bootstrap resampling method \citep[e.g.,][]{Loh2008}. We note that in our analysis only X-ray detected sources are included. Previous similar studies have also included undetected AGN in their measurements, by applying stacking analysis \citep[e.g.,][]{Mullaney2012, Rodighiero2015, Yang2018}. This allows to average the black hole growth rate over long time intervals. However, using only detected sources gives the advantage that the measured X-ray luminosities provide a tracer of black hole luminosity independent of the duty cycle \citep{Carraro2022}. In other words, the quantities we present below trace the average properties of the active population of galaxies at any given epoch.

\subsection{L$_X$ vs M$_*$}
\label{sec_lx_vs_mstar}

Fig. \ref{fig_lx_mstar_all} presents the X-ray luminosity vs. the host galaxy stellar mass, for SB, SF and Q systems (blue squares, green circles and red triangles, respectively), at different redshift intervals, as labeled at the top of each panel. We notice that SB systems showcase an increased median X-ray luminosity (and thus accretion rate)  compared to SF galaxies, and Q galaxies show a deficit of L$_X$ compared to SF systems, at similar M$_*$. This is in agreement with the results of previous similar studies that attribute this behaviour, in conjunction with the SFR-M$_*$ (MS) relation, to a gradual exhaustion of the gas that fuels both the SMBH and the SFR \citep[e.g.,][]{Rodighiero2015}, possibly accompanied by a proportional variation in Eddington ratios \citep[e.g.,][]{Carraro2022}.

Furthermore, L$_X$ presents a mild increase (up to 0.5\,dex) with M$_*$, in particular at high redshifts ($\rm 1.5<z<2.5$). We verified (Appendix \ref{appendix_diff_fields}) that this trend is consistent in all four fields considered in this work. In COSMOS, our average L$_X$-M$_*$ relations generally follow similar patterns, but have lower normalizations \citep[in line with][that used detected X-ray sources in COSMOS]{Carraro2022}. This is because the COSMOS field is characterized by an L$_X$ distribution skewed towards lower values (Fig. \ref{fig_lx_distrib}). Similar trends have also been found in previous studies \citep[e.g.,][]{Mullaney2012, Rodighiero2015, Yang2018, Carraro2020, Carraro2022}, but a more direct comparison can be made with the results of \cite{Carraro2022} that used only the detected X-ray sources of the \cite{Carraro2020} sample. They find a (mild) increase of L$_X$ with M$_*$, similar to our results. 


The increase of L$_X$ with M$_*$ we detect, is most probably driven by the combined effect of the underlying slope of the M$_{BH}$-M$_*$ relation and the variation of the mean Eddington ratio $\lambda\propto$L$_X$/M$_*$ \citep{Carraro2022}. On the assumptions of a relatively constant M$_{BH}$-M$_*$ relation for AGN up to $\rm z \sim 2$ \citep[e.g.,][]{Suh2019}, the gradual flattening and overall decrease of our L$_X$-M$_*$ with cosmic time, could be ascribed to a decrease of the underlying mean Eddington ratio, especially evident for the most massive galaxies allegedly hosting, on average, the most massive SMBHs. Indeed, as we show in Appendix \ref{appendix_lx_dependence} (Fig. \ref{fig_lx_mstar_lxsplit}), the increase in L$_X$ with M$_*$ at high redshift is mostly driven by the most luminous sources, a possible signature that SMBHs in the more massive galaxies tend to accrete at larger rates at earlier epochs.

\subsection{Lx/SFR  vs M$_*$}
\label{sec_lxsfr_vs_mstar}

In Fig. \ref{fig_lxsfr_mstar_all}, we plot the ratio $\rm log(L_X/SFR)$ as a function of M$_*$, for different redshift intervals, as labelled at the top of each panel. Our data suggest that SB have, at all redshifts explored here, a median L$_X$/SFR ratio a factor of $\sim 2-3$ and $\sim 10$ lower than SF and Q galaxies, respectively, with only a weak dependence on stellar mass. Previous studies have found either a positive correlation  \citep[][]{Rodighiero2015, Yang2018, Carraro2020} or a flat relation \citep{Mullaney2012} between $\rm log(L_X/SFR)$ (or BHAR/SFR) and M$_*$, with the caveat that these studies included stacked sources in their analysis. We also note that the amplitude of the $\rm log(L_X/SFR)$-M$_*$ relation in our results is about an order of magnitude higher than the amplitude presented in the aforementioned studies. This is mainly due to the inclusion, in our analysis, of X-ray sources detected in wide fields (eFEDS, XMM-{\it{XXL}}, Bo$\rm \ddot{o}$tes), i.e., our sample includes a larger number of luminous AGN (Fig. \ref{fig_lx_distrib}). Furthermore, application of stacking analysis in these previous studies have allowed them to include very faint sources in their calculations. The different luminosities probed among our study and the previous works mentioned above, affect the amplitude of the $\rm log(L_X/SFR)$-M$_*$ relation, however, we do not expect to affect the overall trends (see Appendix \ref{appendix_lx_dependence}).

To examine if the observed trends in the $\rm log(L_X/SFR)$-M$_*$ relation are driven by L$_X$ or SFR, in Fig. \ref{fig_lx_sfr_check} we plot L$_X$ vs. SFR, for SF, SB and Q AGN host galaxies, in the same redshift intervals as in the previous Figures. The median SFR increases by a large factor (up to 3.5\,dex) across M$_*$, when moving from Q to SB galaxies. The corresponding increase of the mean L$_X$ is significantly smaller ($\sim 0.7$\,dex), at all redshift intervals and for all AGN host galaxy classifications. We thus conclude that the strong increase observed in the L$_X$/SFR ratio from SB to Q at fixed M$_*$ is mostly driven by a significant drop in SFR. 

In Appendix \ref{appendix_lx_dependence}, we explore the $\rm log(L_X/SFR)$ as a function of M$_*$, for luminous and low to moderate L$_X$ AGN. Similar trends are found for both AGN populations (Fig. \ref{fig_lxsfr_mstar_lxsplit}). Luminous AGN have higher $\rm log(L_X/SFR)$ amplitude compared to less luminous X-ray sources, at low redshifts. However, this picture reverses, for AGN hosted by Q galaxies, at the highest redshift interval spanned by our dataset. This behaviour is due to how the L$_X$ and SFR vary as a function of M$_*$ for the different AGN host galaxy classifications and for different L$_X$ and redshifts (for more details see Appendix \ref{appendix_lx_dependence}).

We conclude that our results show that Q systems present the highest amplitude in the  $\rm log(L_X/SFR)$-M$_*$ relation due to their lowest SFRs, while SB galaxies have the smallest $\rm log(L_X/SFR)$ ratio due to their highest SFRs.

\begin{figure*}
\centering
  \includegraphics[width=0.67\columnwidth, height=5.7cm]{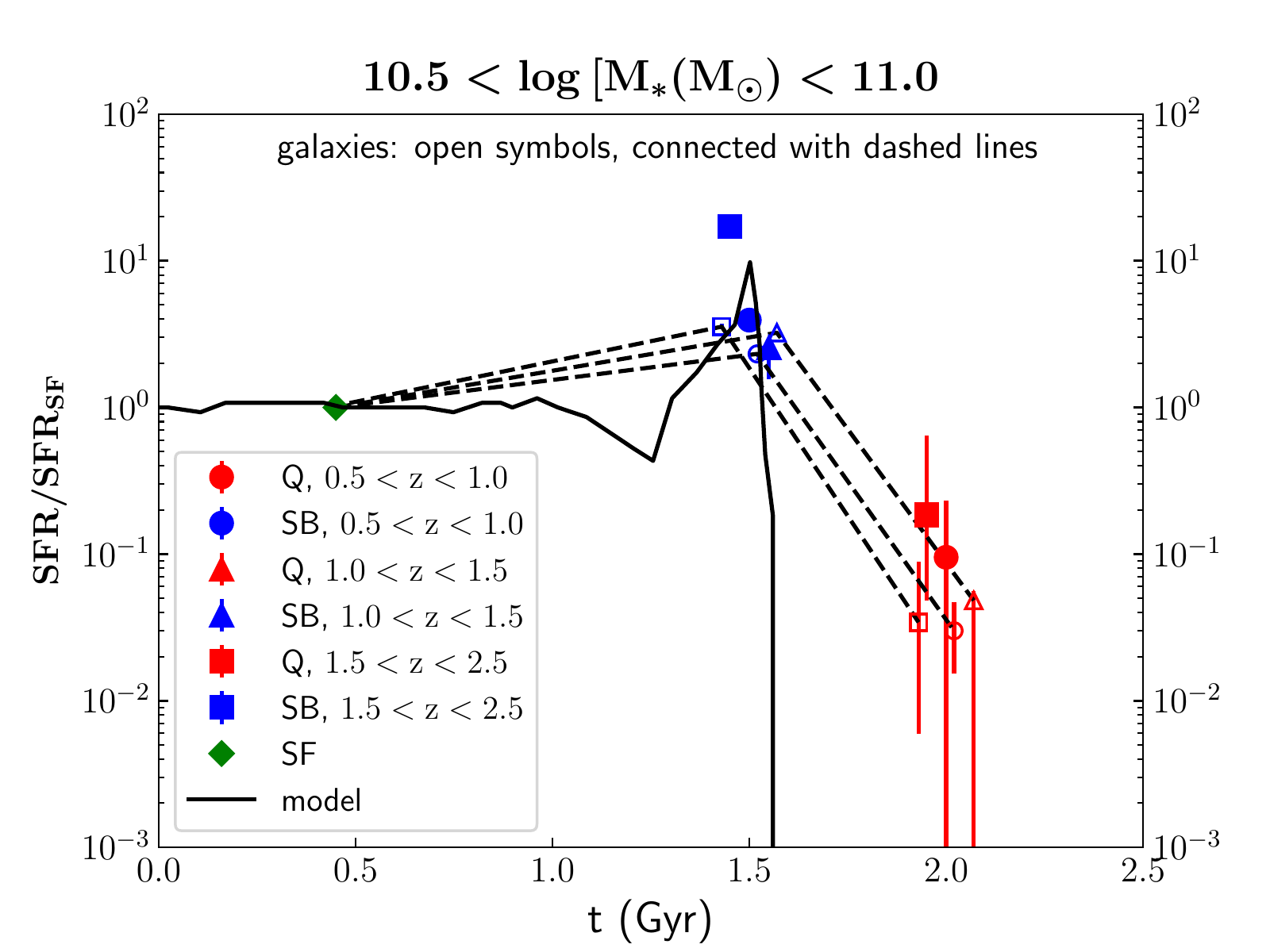}
  \includegraphics[width=0.67\columnwidth, height=5.7cm]{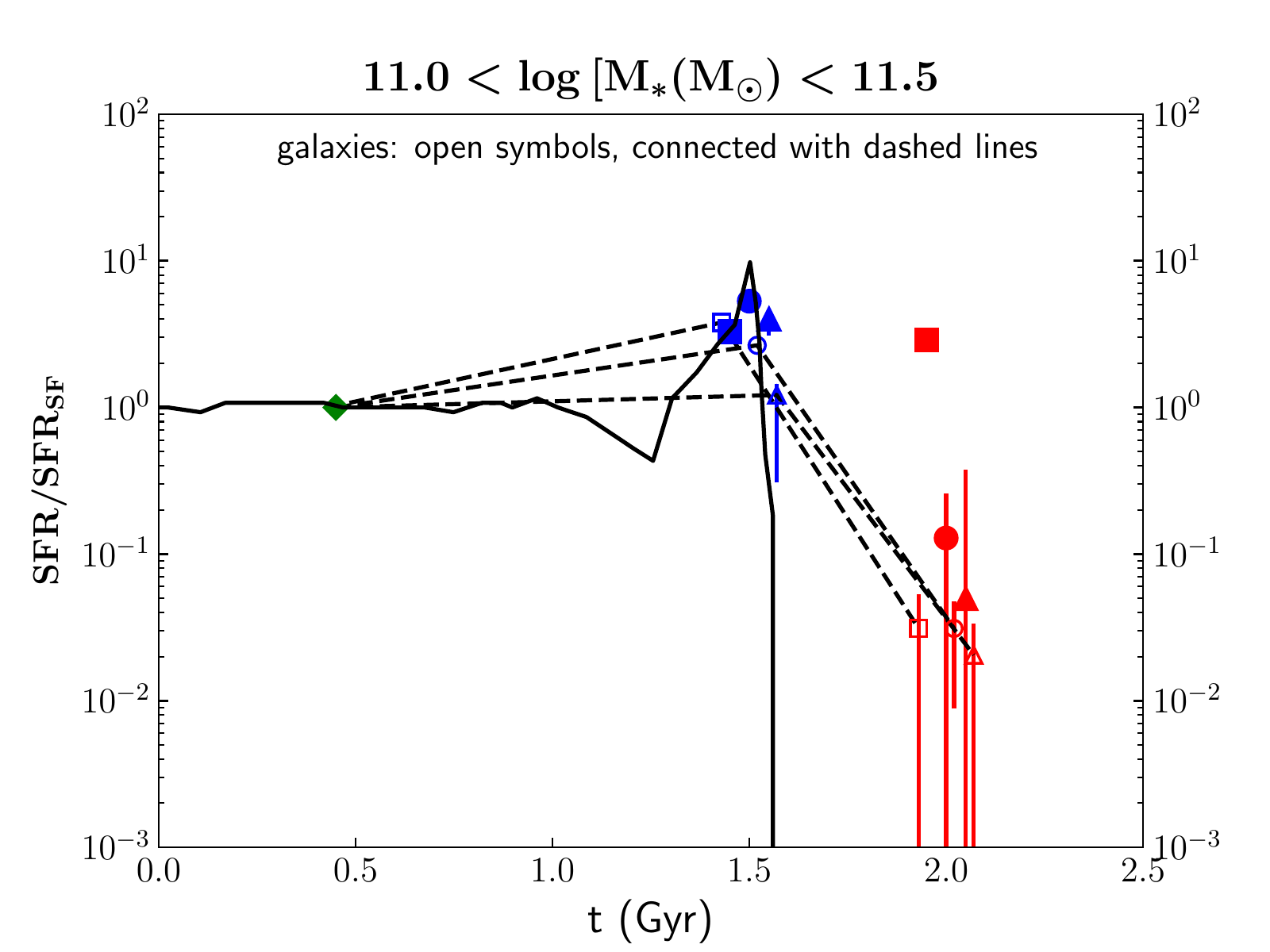}
  \includegraphics[width=0.67\columnwidth, height=5.7cm]{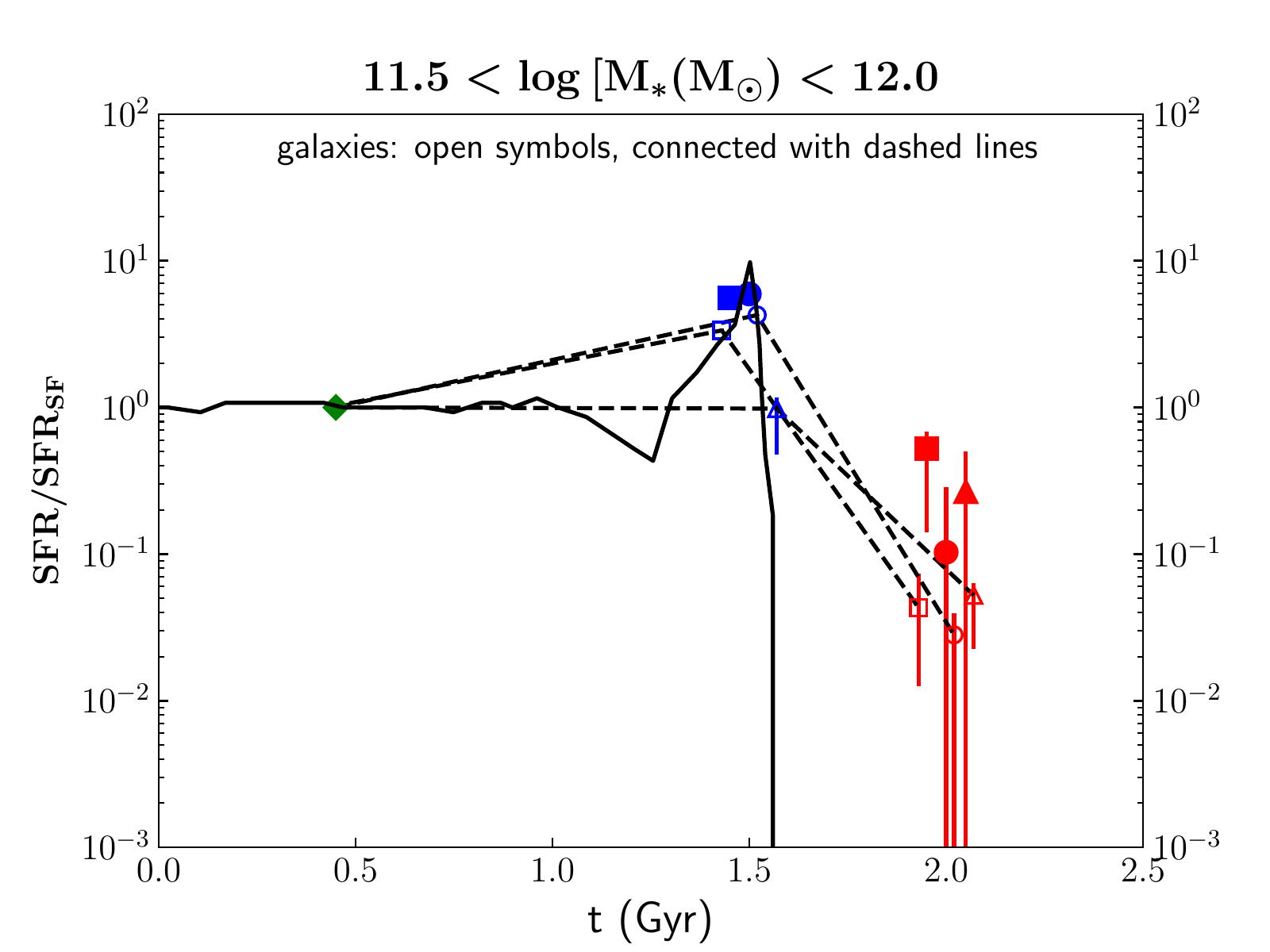}

  \includegraphics[width=0.67\columnwidth, height=5.7cm]{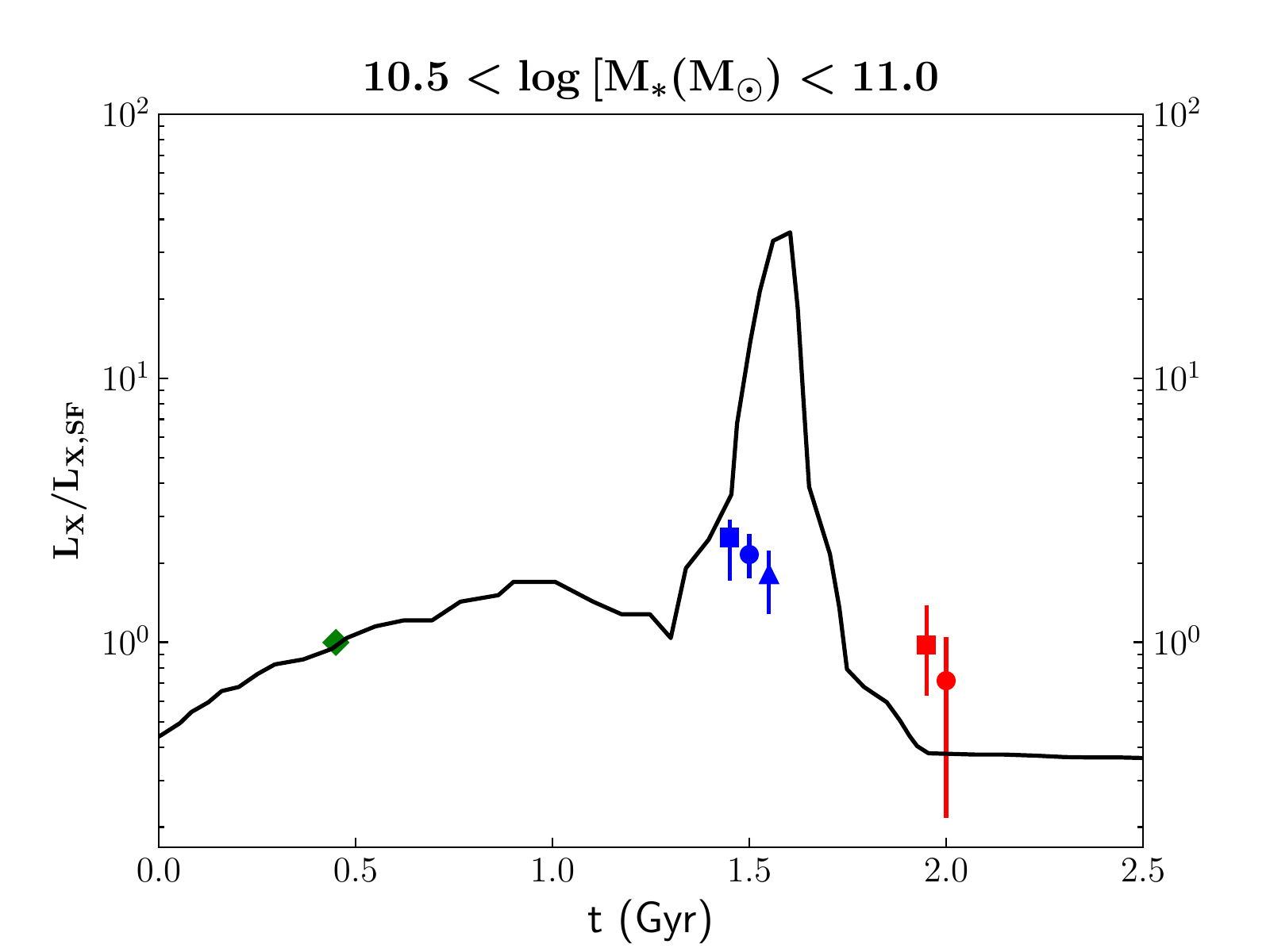} 
  \includegraphics[width=0.67\columnwidth, height=5.7cm]{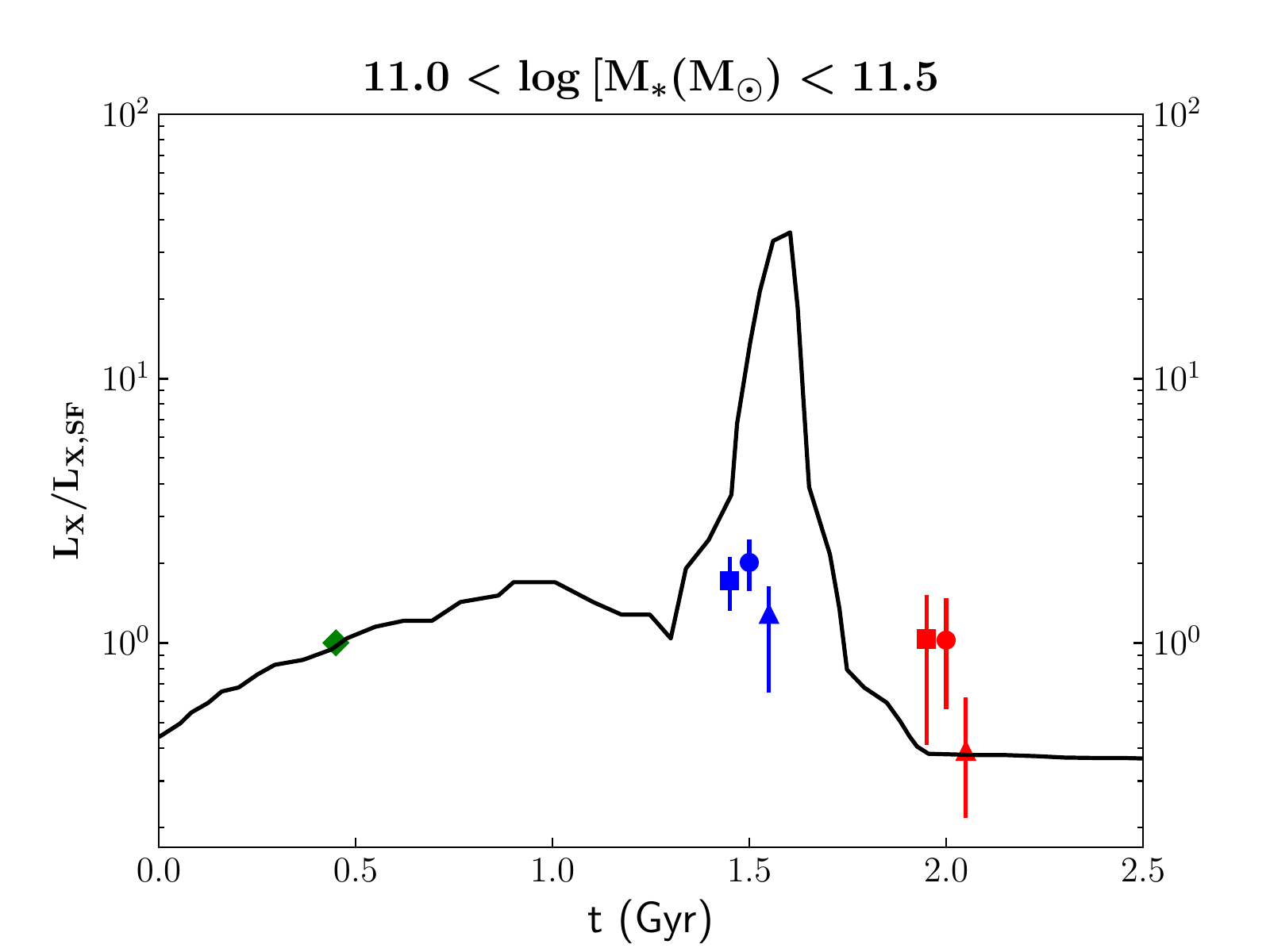} 
  \includegraphics[width=0.67\columnwidth, height=5.7cm]{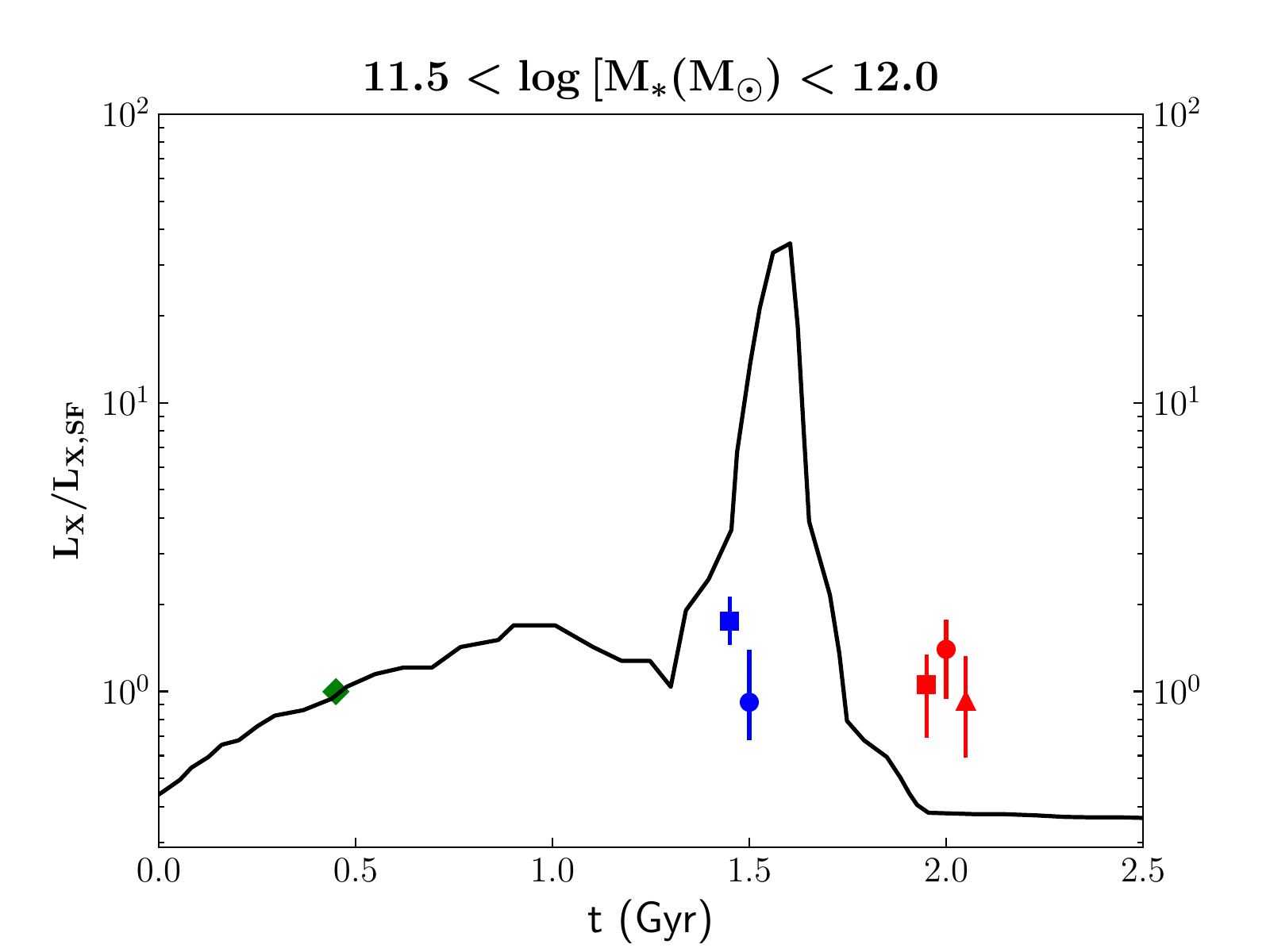} 

  \includegraphics[width=0.67\columnwidth, height=5.7cm]{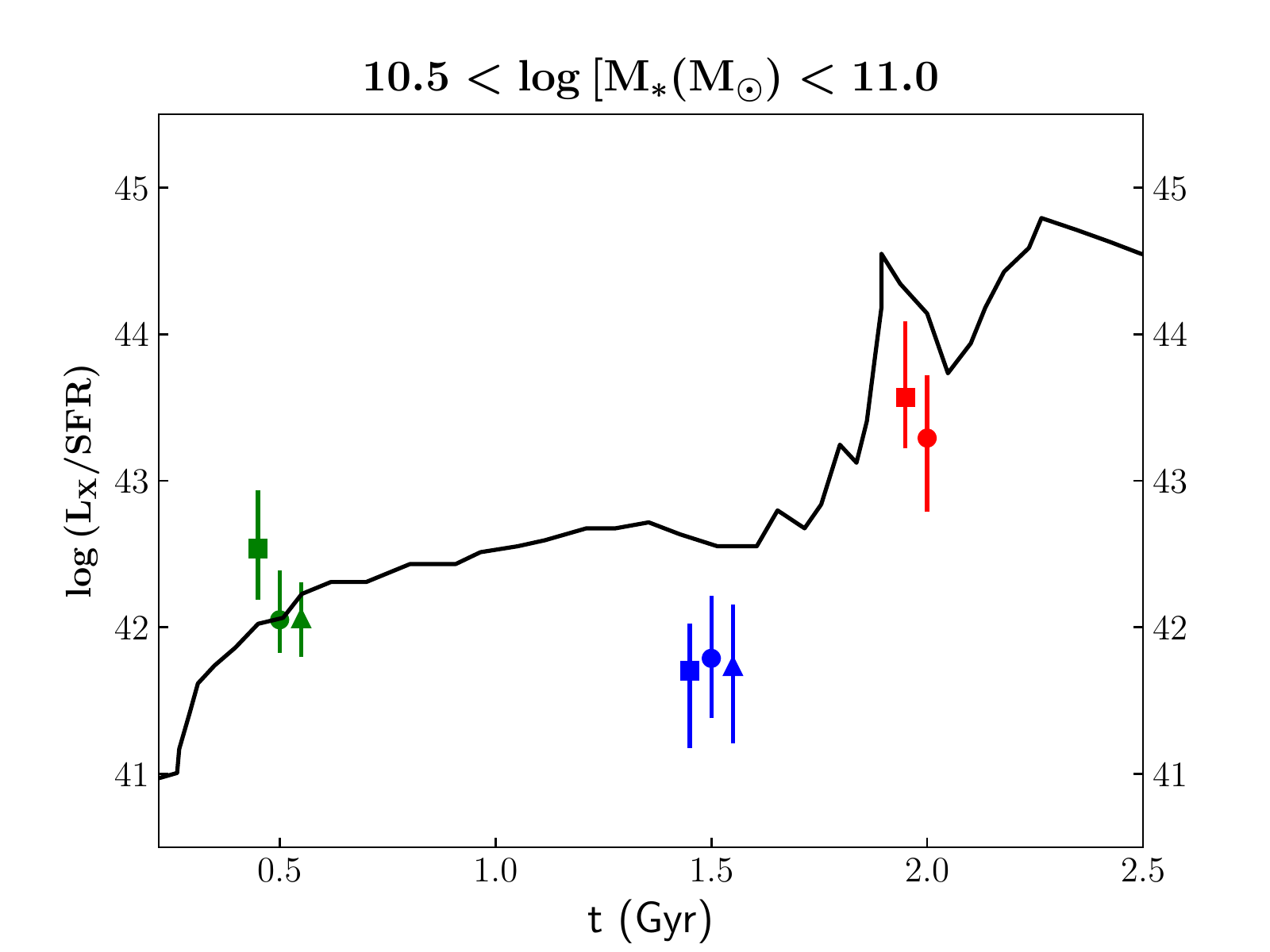}
  \includegraphics[width=0.67\columnwidth, height=5.7cm]{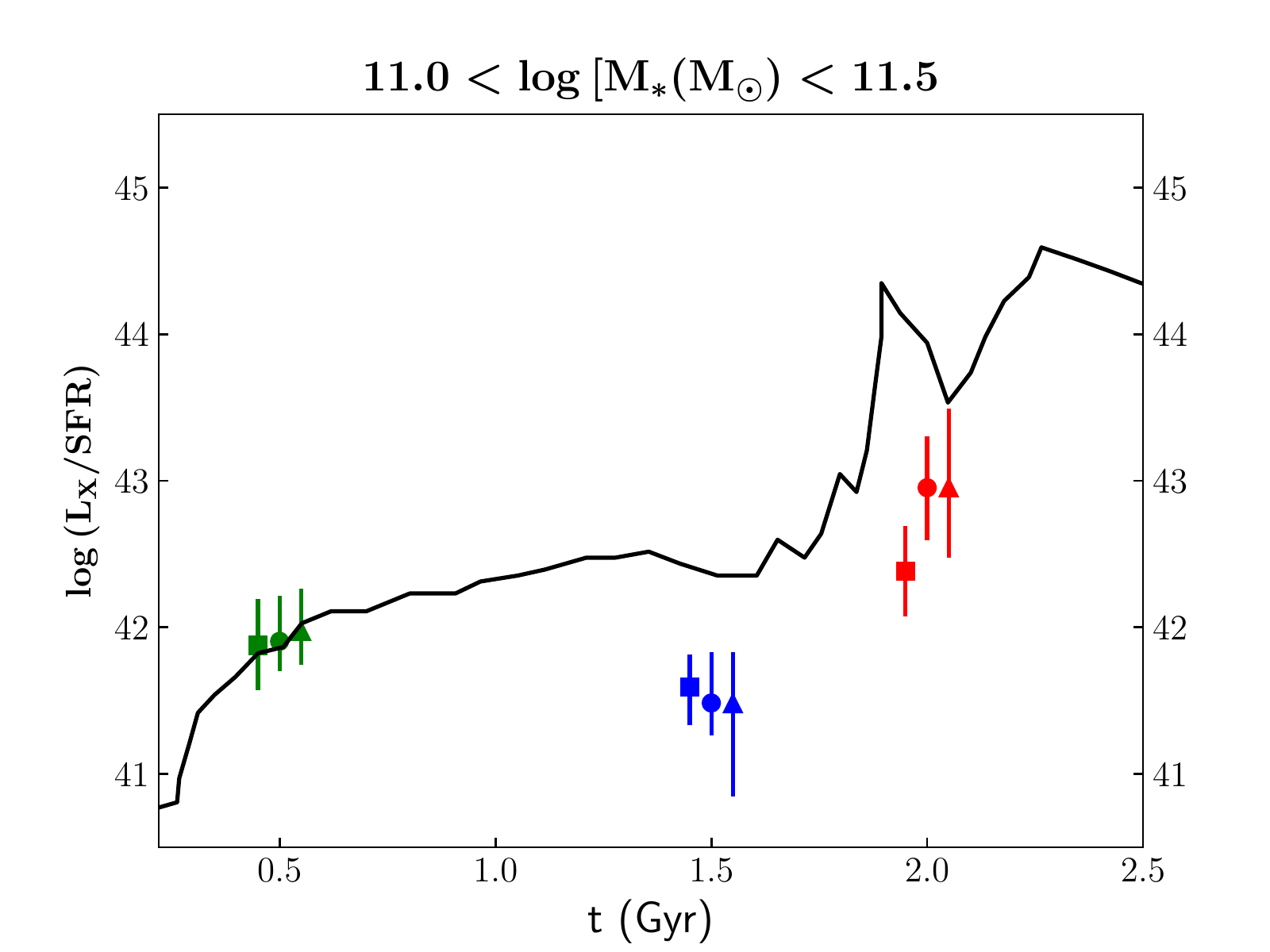}
  \includegraphics[width=0.67\columnwidth, height=5.7cm]{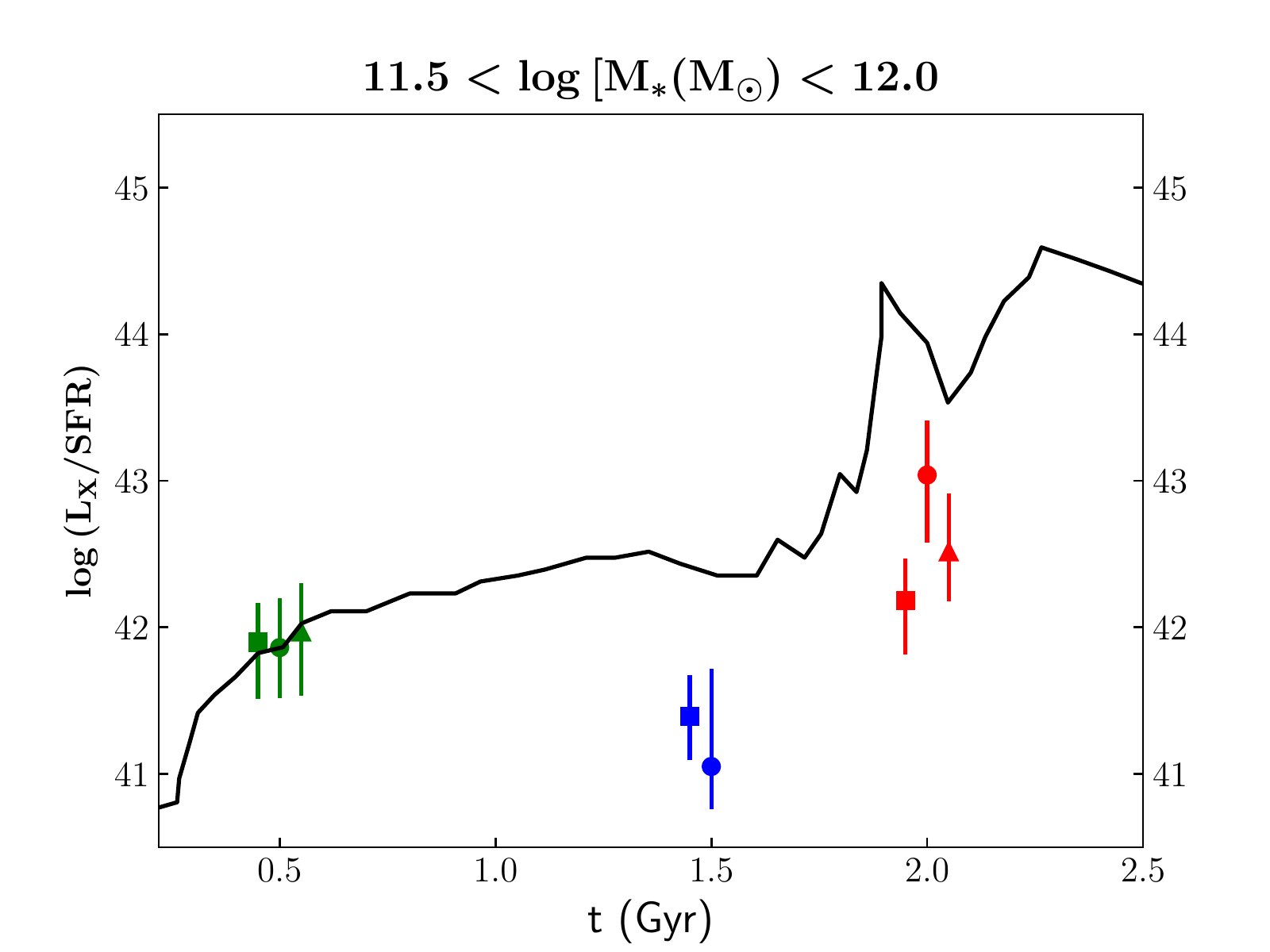}

  \caption{Evolution of SFR (top panels), L$_X$ (middle panels) and their ratio $\rm log L_X/SFR$ (bottom panels) with time. Each row corresponds to a different M$_*$ interval, as indicated as the top of each plot. Filled symbols present the results for galaxies that host X-ray AGN. Open symbols connected with dashed lines show the results for non-AGN galaxies (top panels). The black lines present the predictions of hydrodynamical models that follow the black hole growth and the star formation during a galaxy merger \citep{DiMatteo2005, Hopkins2012}.}
  \label{fig_models_mass}
\end{figure*} 



\section{Discussion}
\label{sec_comp_models}

As anticipated in Sect. 1, studying AGN statistical samples averaged in bins of stellar mass, SFR and L$_X$, provides clues to the overall connection among these properties, but it does not necessarily provide insight into the actual evolutionary phases, if any, between the starburst and star forming/quiescent phases. To extract possible evolutionary sequences characterising AGN galactic hosts, we need to connect the Lx and SFR of SB, SF and Q into a time sequence, and check whether such sequences are comparable to predictions from theoretical models \citep[e.g.,][]{Rodighiero2015}. To this purpose, in this Section, we study how the L$_X$/SFR ratio varies along the different phases of the galaxy, from the SF to the SB and Q phases, and compare our observations with theoretical light curves from SMBH-galaxy evolutionary models.

\subsection{Evolution of L$_X$/SFR at different galaxy evolutionary sequences} 
\label{sec_evolution}

We start by examining how the L$_X$/SFR evolves as the galaxy "transits" from one phase to the other. We assume that the (host) galaxy is initially in a star forming phase. Then, the galaxy merges with another galaxy and the gas is converted to stars in intense bursts of star formation (SB phase), $\sim 1.5$\,Gyr later. Finally, when nearly all the gas has been depleted, the galaxy enters the Q state (at $\sim 2$\,Gyr) \citep[e.g.,][]{DiMatteo2005}. The transition times associated to each phase have been chosen on the assumption that these mirror those suggested by theoretical models, which are, in a typical binary merger scenario at $\rm z\sim 1-2$, $\sim 1$\,Gyr between the SB and SF phases and 0.5\,Gyr between the SF and Q phase \citep[e.g.,][]{DiMatteo2005,Springel2005, Hopkins2012}.

In Fig. \ref{fig_models_mass}, the top, middle and bottom panels show, respectively, the putative evolution of the SFR, L$_X$ and the L$_X$/SFR ratio, when the galaxy "transits" from the SF to the SB and then to the Q phase, assuming that these are correlated to each other. The results for different M$_*$ selections are presented, as indicated in the top of each panel. We do not show results for the lowest stellar mass bin ($\rm 10<log\,[M_*(M_\odot)]<10.5$), since there are not enough Q AGN host galaxies in this M$_*$ bin in our sample to allow robust calculations. The estimated median quantities reported are largely independent of the exact cuts in redshift and/or stellar mass interval chosen, and still hold true when assuming some evolution in the host galaxy stellar mass, following semi-empirical models, as discussed in the next section. We start discussing the bottom panels of Fig. \ref{fig_models_mass}, where we show that the L$_X$/SFR tends to gradually increase by up to $\sim 1-1.5$\,dex when the host galaxy becomes Q.

To examine what drives the variation of the L$_X$/SFR with time, in the top and middle panels of Fig. \ref{fig_models_mass}, we normalize the SFR and the L$_X$, respectively, to the SFR and L$_X$ of the SF (main sequence) galaxies and study them as a function of time. Both SFR and L$_X$ increase as the galaxy moves from the SF to the SB and then decrease in the Q state. We find that the SFR increases by nearly an order of magnitude in the transition from the SF to the SB stage and then drops by almost two orders of magnitude from the SB to the Q phase. On the other hand, L$_X$ increases by, only, up to a factor of two in the SB phase and then, in the Q phase, drops to similar values with those observed for SF galaxies. These results suggest that the L$_X$/SFR fluctuations we observe during the different galaxy evolutionary sequences are largely driven by variations in the SFR. It is also interesting to note that the aforementioned fluctuations of SFR and L$_X$ are nearly independent  of both redshift and galaxy stellar mass.

Our next step is to compare our inferred SFR evolutionary tracks with those from galaxies not detected as AGN. \cite{Mountrichas2021c, Mountrichas2022a, Mountrichas2022b} constructed galaxy control samples  applying the same photometric requirements and SED fitting techniques (same templates and parameter values) as in the X-ray sources adopted in this work. Non-X-ray AGN systems have also been excluded from the control sample \citep[for more details see Sect. 3.3 in e.g.,][]{Mountrichas2022b}. We, thus, combine the galaxies from the Mountrichas et al. datasets and compile a sample of $\sim 130,000$ sources within the mass completeness limits, and similarly classify these galaxies into SB, SF and Q, following the same method as for the X-ray AGN (see Sect. \ref{sec_classification}). We then normalize the SFR of each class based on the SFR of the SF galaxies. The results are shown in the top panels of Fig. \ref{fig_models_mass} (open symbols connected with dashed lines), for different M$_*$ regimes. We find that for non-AGN sources the SFR increases in the SB phase and drops in the Q stage in a similar manner as for the X-ray AGN. The similarity in the variation of SFR from SB to Q galaxies between AGN and non-AGN galaxies, suggest that the AGN has a negligible impact on the quenching. Theoretical studies have suggested that the apparent lack of a (inverse) correlation between AGN activity and SFR found in observational works, could be induced by the different timescales characterizing the
two processes \citep[e.g., ][]{Hickox2014, Ward2022}. Theoretical models indeed suggest that a delay of a few millions of years is required for the impact of a single episode of nuclear activity to have a measurable effect on star formation. The presence of such delays may explain, at least in part, why all galaxies of similar stellar mass in our sample share similar drops of the SFR from the SB/SF to the Q state, irrespective of their AGN activity. In other words, the combination of a delay and of relative short AGN lifetimes \citep[e.g.,][]{Martini2003, Shankar2004}, could both contribute to wash out causal links between AGN luminosities and host galaxies SFR.

Another viable solution could be to start from a SB phase, on the assumption that the galaxy is forming out of the cooling of pristine gas clouds, and then gradually converts into a SF and finally into a Q phase \citep[e.g.,][]{Granato2004, Granato2006, Lapi2006,Lapi2018}. Although, without clear age and structural discriminators, our current data cannot safely distinguish between these two evolutionary routes, our measurements presented in Fig. \ref{fig_lx_mstar_all}, suggest that SB could be the starting phase in a galaxy's lifetime. 

\begin{figure}
\centering
  \includegraphics[width=0.75\columnwidth, height=5.7cm]{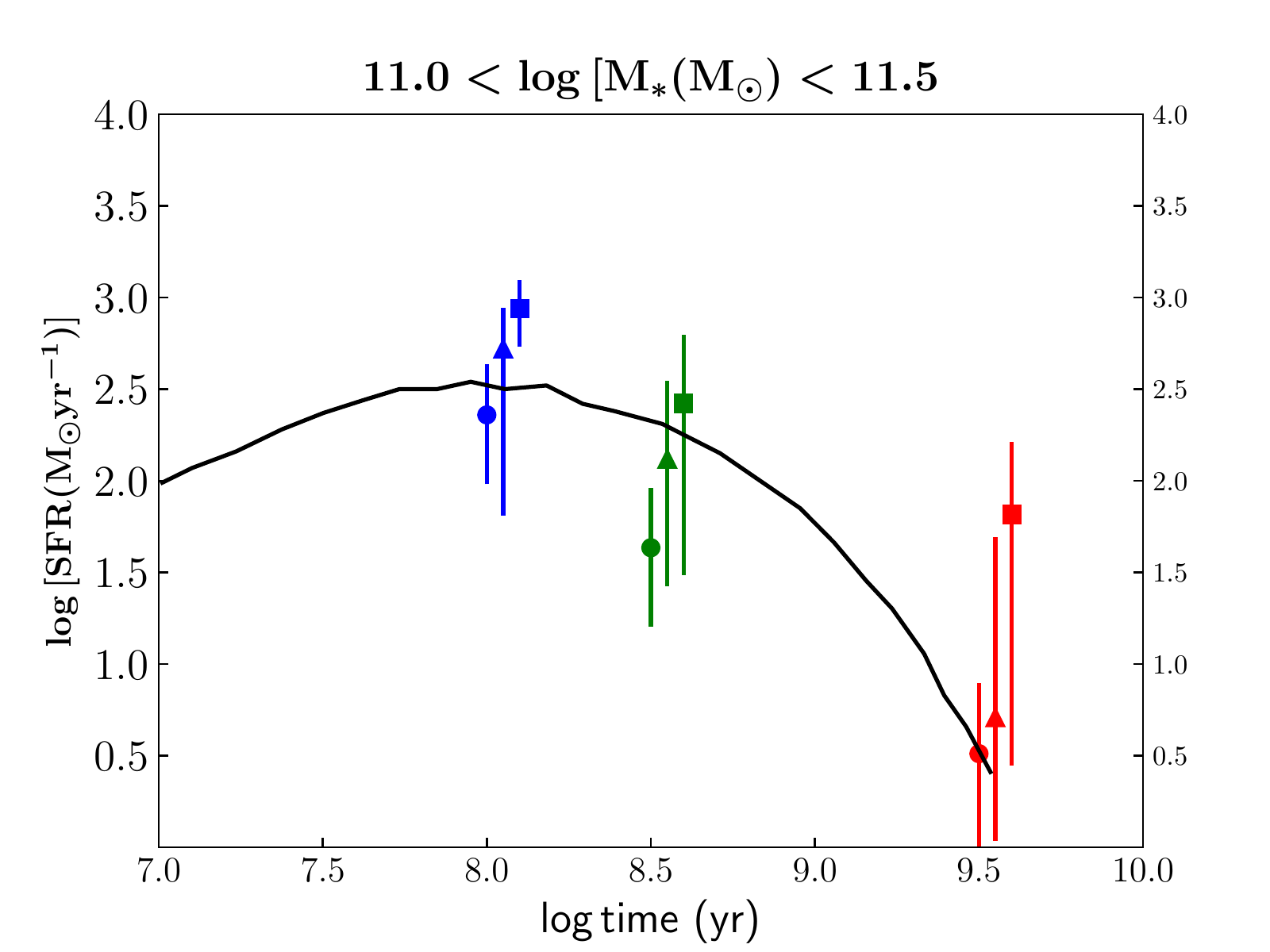}
  \includegraphics[width=0.75\columnwidth, height=5.7cm]{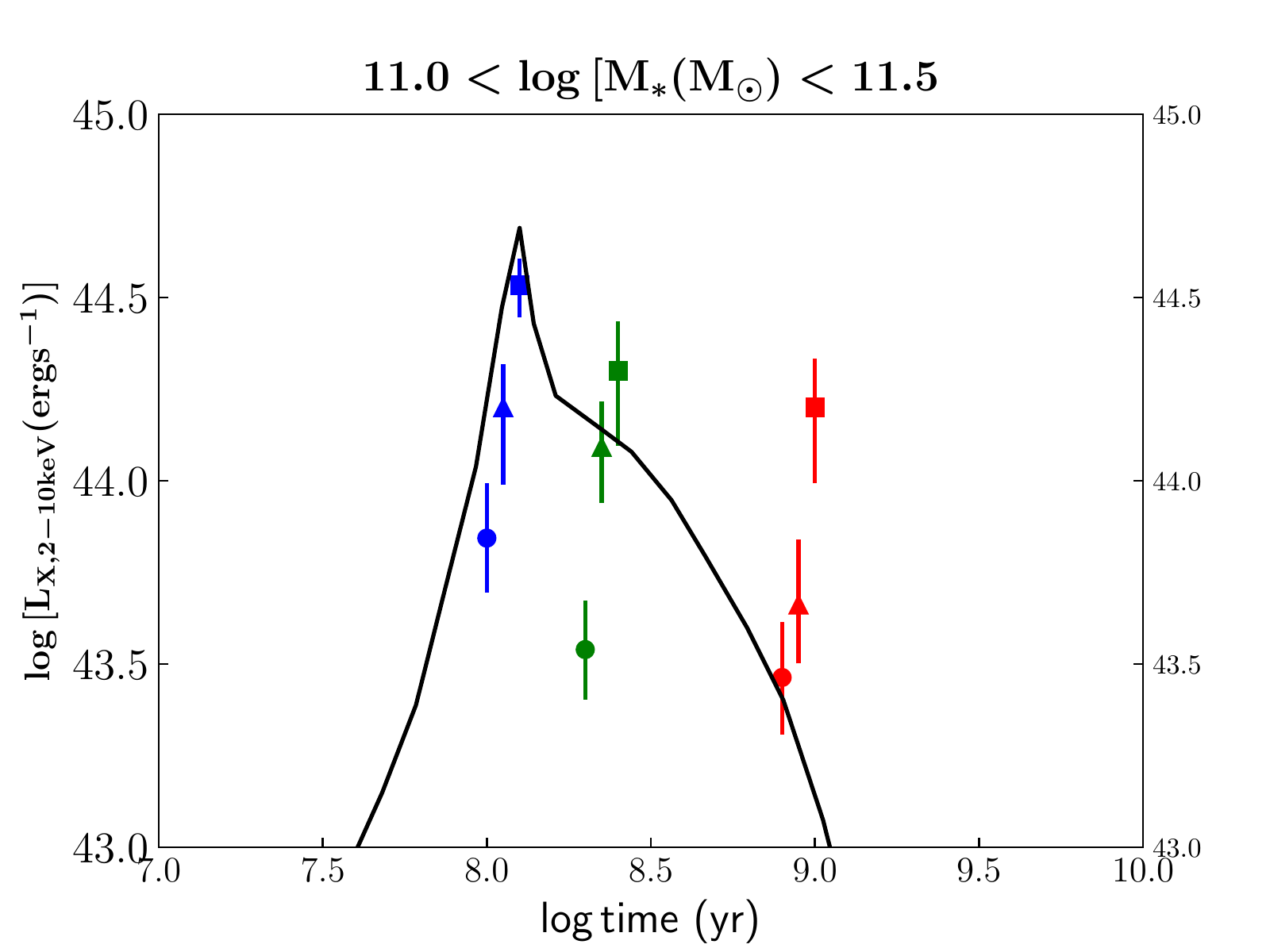}
  \includegraphics[width=0.75\columnwidth, height=5.7cm]{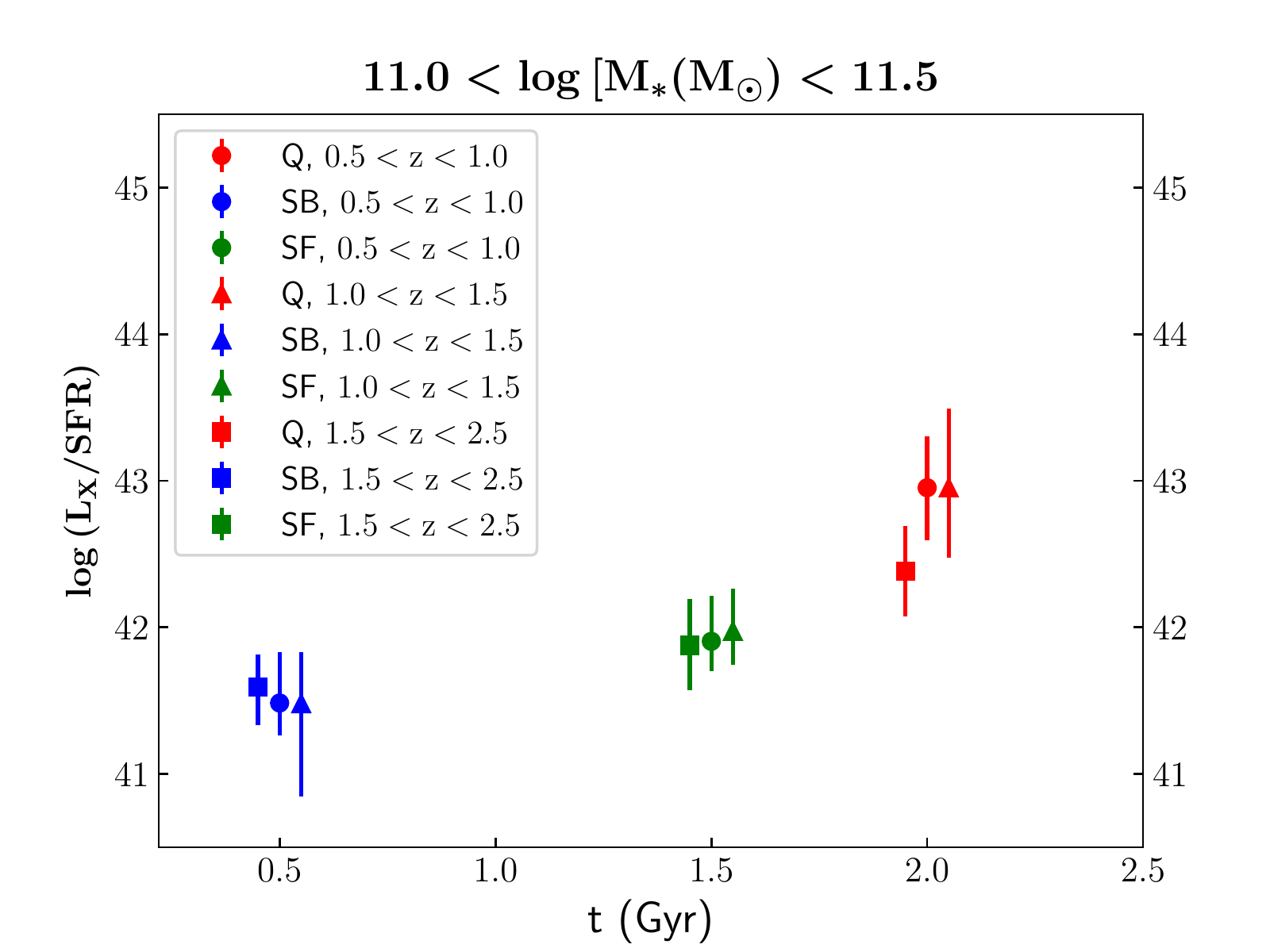}
  \caption{Evolution of SFR (top panel), L$_X$ (middle panel) and their ratio $\rm log L_X/SFR$ (bottom panel) with time, for AGN host galaxies with $\rm 11.0<log\,[M_*(M_\odot)]<11.5$. In this exercise, we assume that the galaxy starts its life in a SB phase and then it enters the SF and finally becomes Q.  The black lines show the predictions of the model presented in Granato et al. 2004. The x-axis in the top and middle panels, present the galactic age (in log), as shown in Figures 2 and 3 in Granato et al. 2004.}
  \label{fig_models_mass_sb}
\end{figure}

In Fig. \ref{fig_models_mass_sb}, we plot the SFR (top panel), the L$_X$ (middle panel) and their ratio L$_x$/SFR (bottom panel), as a function of time, for AGN host galaxies with $\rm 11.0<log\,[M_*(M_\odot)]<11.5$, assuming that the galaxy starts its life in a SB phase. The x-axis in the top and middle panels present the galactic age (in log) to allow us to compare our observational measurements with the theoretical model of \cite{Granato2004} (see next Section). Similarly to the results presented above, the L$_X$/SFR ratio increases by up to $\sim 1-1.5$\,dex when the host galaxy becomes Q. Both the SFR and L$_X$ drop as the galaxy evolves from the SB to the SF and then the Q phase. This decrease is higher for SFR ($\sim 2$\,dex) compared to L$_X$ ($\sim 1$\,dex). Similar results are found for the other stellar mass bins.

\subsection{Comparison with theoretical predictions}
\label{sec_LC}


To set the stage between our data and theoretical models, we follow \cite{Rodighiero2015} and compare in Fig. \ref{fig_models_mass} our empirical renditions of AGN light curves with those extracted from hydrodynamic simulations inclusive of AGN feedback \citep[][]{DiMatteo2005, Hopkins2012} We take the typical light curve of a galaxy hosted in a dark matter halo of 160 km/s \citep[black lines in Fig. 3 of][]{Rodighiero2015} \footnote{Our results would be similar for other choices of dark matter haloes. As shown by \cite{Rodighiero2015}, in fact, light curves in larger haloes are predicted to be very similar.}. The empirical light curves suggest that the AGN luminosity increases only by a factor of a few from the SF to the SB phase, and a similar drop is observed in the Q phase. This behaviour is not entirely aligned with what predicted by models which suggest that during the growth phase the BH grows by a factor of $>100-1000$ depending on the assumed initial BH seed. We also checked, via Monte Carlo simulations, that typical random measurement errors of 0.5 Gyr in galaxy ages and of 0.15 dex in X-ray luminosities, are not capable to wash out the peak of the AGN light curve and to reconcile data with models, thus suggesting that this discrepancy we observe between predicted and observed AGN X-ray light curves is robust. 

On the other hand, the SFR is observed to actually behave as predicted, with an increase of a factor of a few up to an order of magnitude, followed by a drop of two orders of magnitudes, similar to what expected in AGN feedback models where an AGN wind is expelling mass from the host galaxy or at least injecting energy/momentum thus decreasing the possibility for any further gas fragmentation.

In Fig. \ref{fig_models_mass_sb}, we assume instead that a galaxy starts its life in a SB phase and compare our observational measurements with the predictions of the model presented in \cite{Granato2004}. We find even in this case that the SFR evolves from the SB to the SF and then to the Q phase, as predicted (top panel of Fig. \ref{fig_models_mass_sb}). However, L$_X$ (middle panel of Fig. \ref{fig_models_mass_sb}) drops by $\sim 0.5$\,dex, compared to 1.5\,dex predicted by the theoretical model. However, including observational uncertainties may improve the match to the models in this second type of models, at least at intermediate redshifts.

Taking these results at face value, we would conclude that in order to achieve a condition where the AGN luminosity drops only by a factor of a few from the peak whilst the SFR decreases significantly, a high AGN feedback efficiency and/or coupling with the interstellar medium (ISM) should be characterising AGN feedback processes. 

The relatively limited pre-peak increase in the AGN luminosity, and to a lesser extent in our SFRs, observed in our data, may signal the fact that in our samples the SF and SB phases are close to each other. In addition, if our simulations inclusive of observational errors, are not able to resolve the tension in the discrepancy between predicted and observed AGN light curves, at least for some models \citep[e.g.,][]{Springel2005, Hopkins2006}, then it would imply that the post-peak phase of the AGN light curve does not decrease as sharply as predicted by some models, and this is in fact what seen by e.g., \cite{Lapi2014}  to reproduce their FIR galaxy light curves, and also by some continuity equation models that require extended light curves \citep[e.g.,][]{Shankar2013, Aversa2015}. 

Interestingly, we find that for each of our galaxy classes, the increase in X-ray AGN luminosity is always smaller than the one in SFR within the same time lag, and in the SB the X-ray luminosity is nearly constant, despite the SFR still increasing by nearly an order of magnitude. Again, this behaviour in our data is not entirely paralleled especially in the merger models, which could either be a consequence of selection effect and/or observational errors, but also it could reflect the fact that the BH has reached a self-regulated state which prevents further substantial growth. 

We note that all our results discussed above remain valid under the assumption that the stellar mass does not grow more than a factor of $\sim 2$ within the timeframe considered, i.e., $\sim 2$\,Gyr, or even by following back in time the stellar progenitors. We checked in fact that by selecting the SF and SB progenitors of the quiescent galaxies of a given chosen M$_*$ following, e.g., the average stellar mass growth histories of \cite{Moster2018}, or even \cite{Grylls2019, Behroozi2019} \citep[see Fig. 3 in][]{Shankar2020}, we find very similar results for the L$_X$, SFR evolution with time as the ones reported in Fig. \ref{fig_models_mass}.

\section{Conclusions}
\label{sec_summary}

We used $\sim 5500$ X-ray AGN detected in four fields, namely the {\it{COSMOS-Legacy}}, the Bo$\rm \ddot{o}$tes, the XMM-{\it{XXL}} and the eFEDS fields. The X-ray sources span a redshift range of $\rm 0.5<z<2.5$ and about three orders of magnitude in X-ray luminosity ($42.0 < \rm log\,[L_{X,2-10keV}(ergs^{-1})] < 45$), while their stellar mass ranges from $10.0<\rm log\,[M_*(M_\odot)]<12.0$. These sources meet strict photometric selection requirements and various selection criteria have been applied to ensure that only sources with robust host galaxy measurements are included in the analysis. The latter have been calculated via SED fitting, using the CIGALE code. Furthermore, our final sample consists only of AGN that satisfy the mass completeness limits of the field they belong. The X-ray sources are then classified into three classes, based on the star formation activity of the host galaxy, i.e., SF, SB and Q. For the classification, the sSFR of each system was used. Our final X-ray sample, consists of 3575 SF galaxies, 939 SB and 917 Q systems. The fraction of each class depends on the redshift, but at all redshift ranges used in our analysis, the bulk of the black hole accretion occurs in SF systems. Our main results can be summarised as follows:

\begin{itemize}

      \item[$\bullet$] We found that SB systems have increased AGN accretion (L$_X$) compared to SF systems, at similar M$_*$. Q galaxies present the lowest L$_X$ among the three classifications. We also find a mild increase of L$_X$ with M$_*$, possibly related to a decrease of Eddington ratio with increasing stellar mass and at fixed M$_{BH}$-M$_*$ slope \citep[e.g.,][]{Carraro2022}. Our results also show a gradual flattening and overall decrease with cosmic time of the L$_X$ at fixed M$_*$, which suggests a decrease of the underlying mean Eddington ratio, in particular for the most massive galaxies. 


    \item[$\bullet$] The amplitude of the L$_X$/SFR ratio is higher for Q systems compared to SF and SB galaxies. We also find that the L$_X$/SFR-M$_*$ relation is nearly flat, in particular for the most massive galaxies. 




    \item[$\bullet$] The ratio of L$_X$/SFR decreases by $\sim 0.5$\,dex when the galaxy enters the SB phase and then increases by almost an order of magnitude in the Q phase. This variation is mostly driven by variations in the SFR, whilst the L$_X$ remains roughly constant from the SF to the SB and the Q states. 
    
    \item[$\bullet$] The fluctuation of SFR is consistent with what predicted by theoretical models, whereas the behaviour of L$_X$ is not in line with merger models that predict an increase of L$_X$ by a factor of $>100-1000$ in the SB phase. 
    
    \item[$\bullet$] We also study the evolution of SFR for a galaxy control sample of non-AGN systems and found that it is very similar to that of X-ray AGN. This similarity in the variation of SFR between the two populations may suggest that AGN have a negligible impact on the star formation quenching or that the two processes proceed on different timescales.

\end{itemize}

\section*{Acknowledgements}

GM acknowledges support by the Agencia Estatal de Investigación, Unidad de Excelencia María de Maeztu, ref. MDM-2017-0765. FS acknowledges partial support from the European Union's Horizon2020 research and innovation programme under the Marie Sk\l odowska-Curie grant agreement No. 860744.

\section*{Data Availability}

The data underlying this article will be shared on reasonable request to the corresponding author.



\bibliographystyle{mnras}
\bibliography{mnras_template}{} 




\appendix

\section{L$_X$ vs M$_*$ in different fields}
\label{appendix_diff_fields}

In this Section, we examine the L$_X$-M$_*$ relation separately in each one of the fields, used in our analysis. Our goal is to check whether the measurements in each field are compatible with each other. For this exercise, we have used sources at $\rm 0.5<z<1.0$ since in this redshift range our datasets have the largest number of available sources (Table \ref{table_data}).

\begin{figure}
\centering
  \includegraphics[width=0.8\columnwidth, height=6cm]{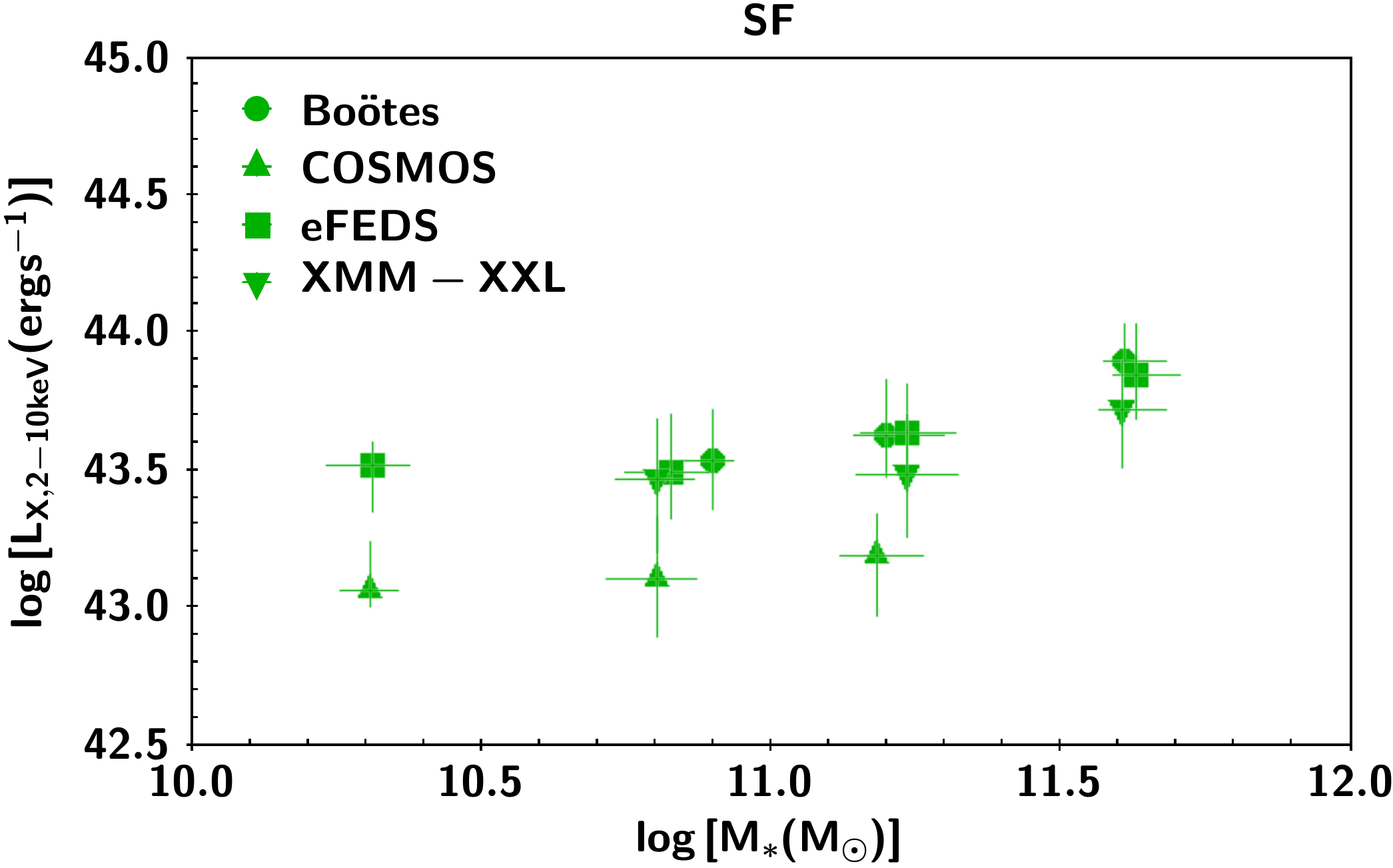} 
  \includegraphics[width=0.8\columnwidth, height=6cm]{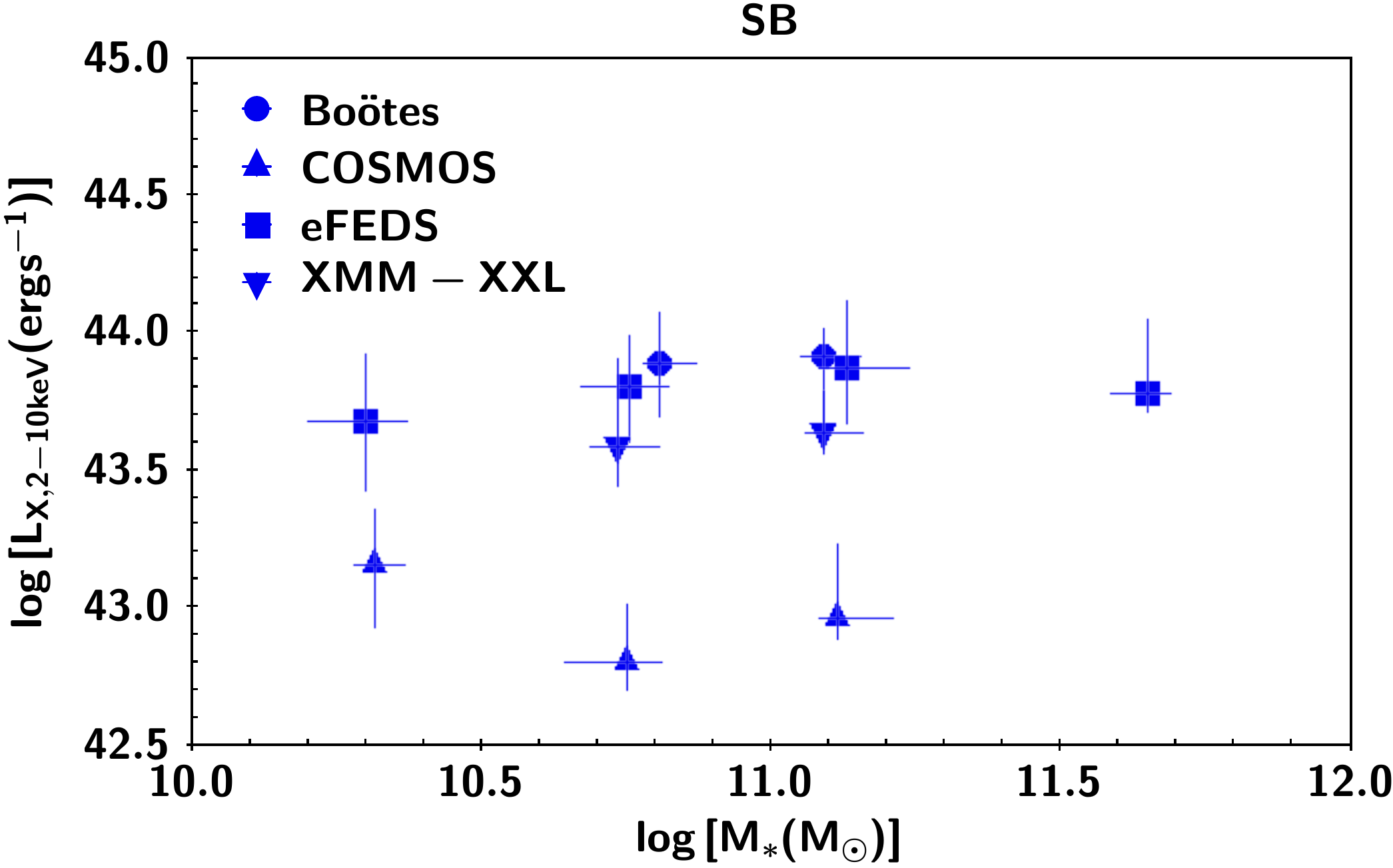} 
  \includegraphics[width=0.8\columnwidth, height=6cm]{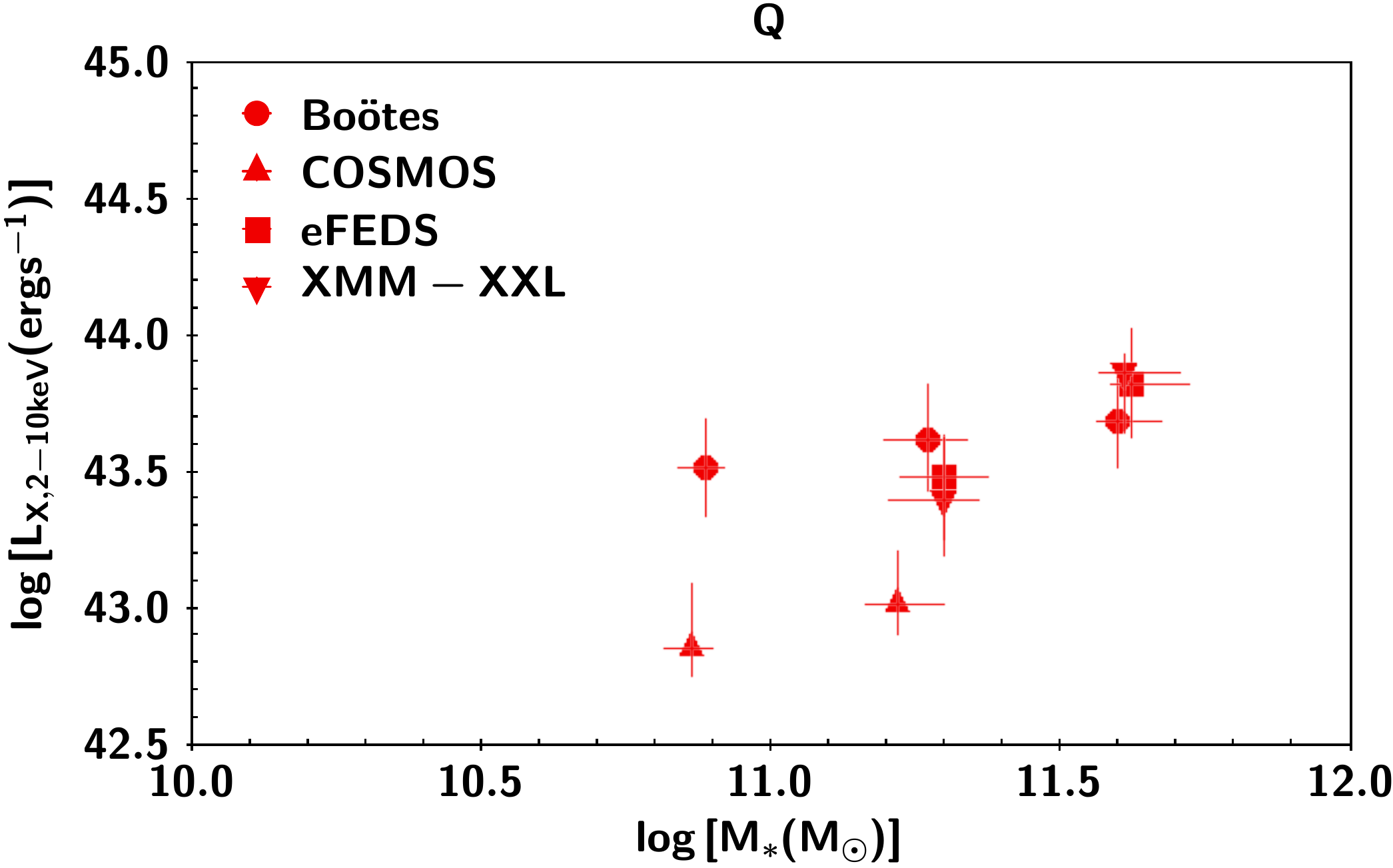}
  
  \caption{L$_X$ vs M$_*$ at each field separately, for sources with $\rm 0.5<z<1.0$. All fields give consistent results within the same redshift, L$_X$ and M$_*$ bins. COSMOS L$_X$ measurements are by $\sim 0.5$\,dex lower, as expected from the L$_X$ distribution of AGN in this field. Most importantly, the observed trends are similar between the four fields.}
  \label{fig_lx_mstar_separate_fields}
\end{figure}

In Fig. \ref{fig_lx_mstar_separate_fields}, we present the L$_X$ as a function of M$_*$ for the Bo$\rm \ddot{o}$tes, COSMOS, eFEDS and XMM-{\it{XXL}} fields, for AGN hosted by SF, SB and Q galaxies. Based on the results, the measurements are consistent among the various field. COSMOS has lower L$_X$ values, for all three classifications, but this is expected since the vast majority of X-ray AGN in this field have low to moderate luminosity sources ($\rm log\,[L_{X,2-10keV}(ergs^{-1})]<44$, Fig. \ref{fig_lx_distrib}). Most importantly, the trends observed are similar in all four fields. Therefore, we conclude that the four fields give consistent results with each other.

\section{Dependence on L$_X$}
\label{appendix_lx_dependence}

In this Section, we examine if the observed trends of L$_X$ and the ratio L$_X$/SFR as a function of M$_*$, presented in Sections \ref{sec_lx_vs_mstar} and \ref{sec_lxsfr_vs_mstar}, are different between low to moderate L$_X$ ($\rm log\,[L_{X,2-10keV}(ergs^{-1})]<44$) and high L$_X$ AGN ($\rm log\,[L_{X,2-10keV}(ergs^{-1})]>44$). 

\begin{figure}
\centering
  \includegraphics[width=0.8\columnwidth, height=6cm]{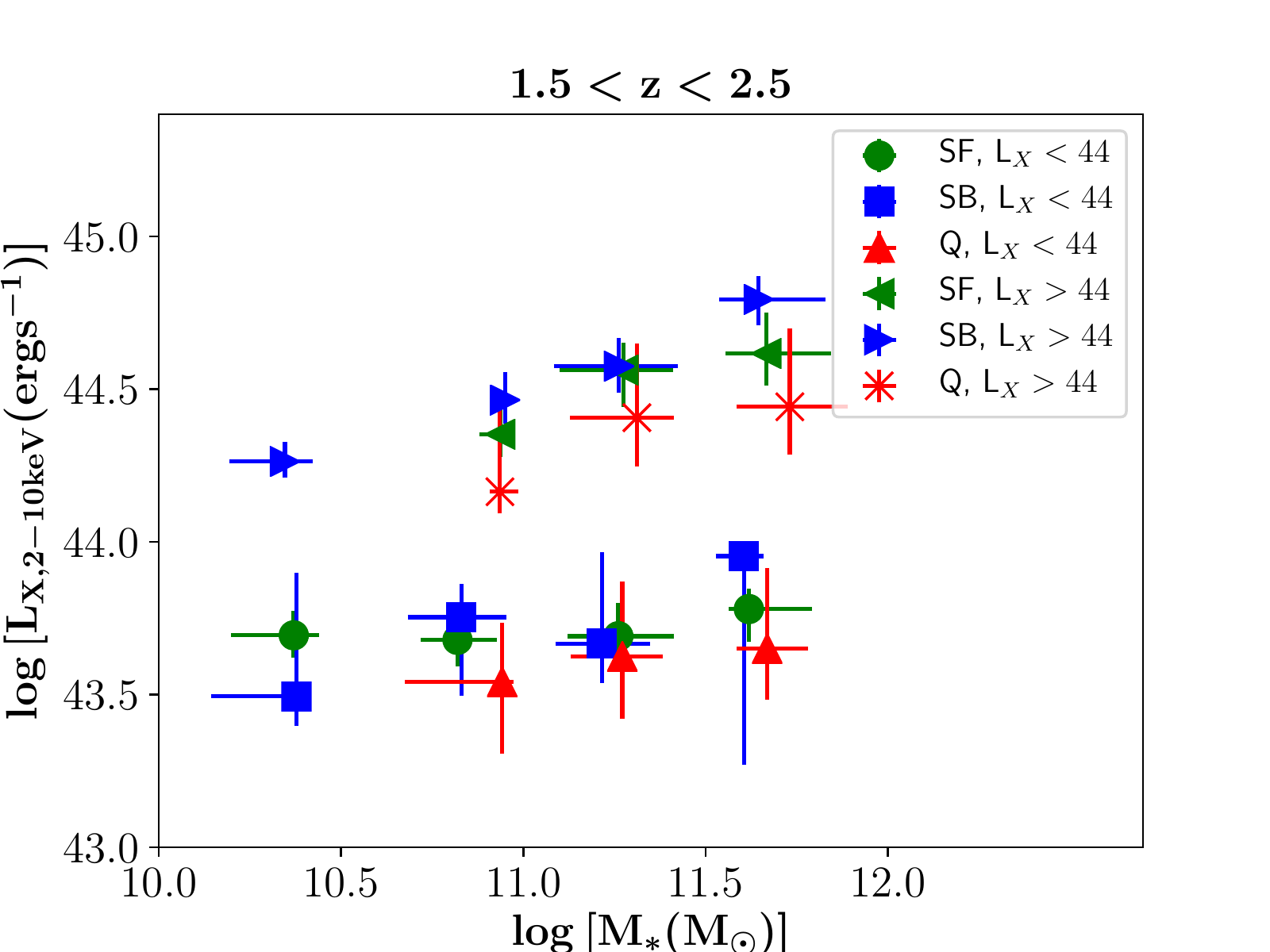} 
  \includegraphics[width=0.8\columnwidth, height=6cm]{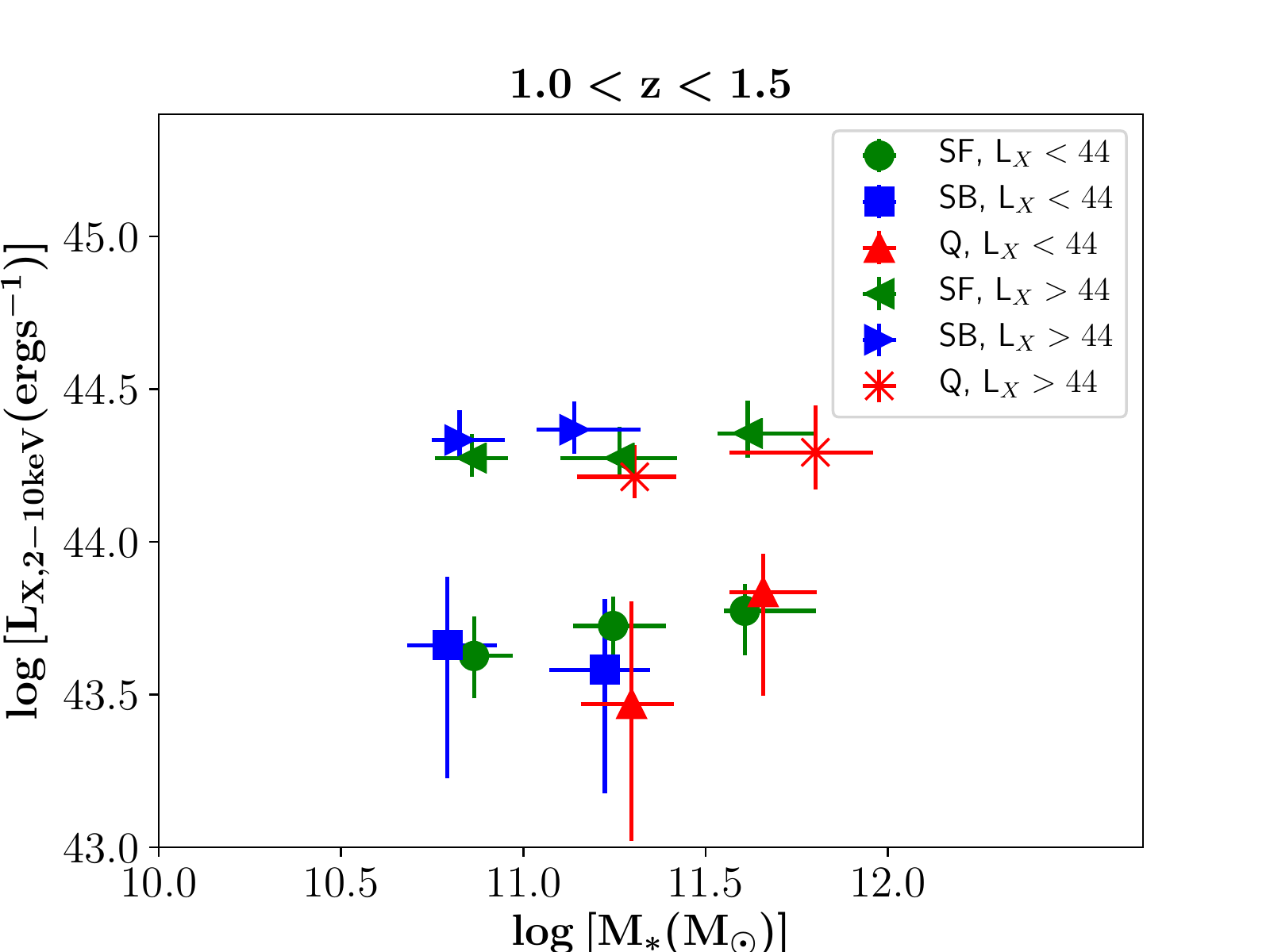} 
  \includegraphics[width=0.8\columnwidth, height=6cm]{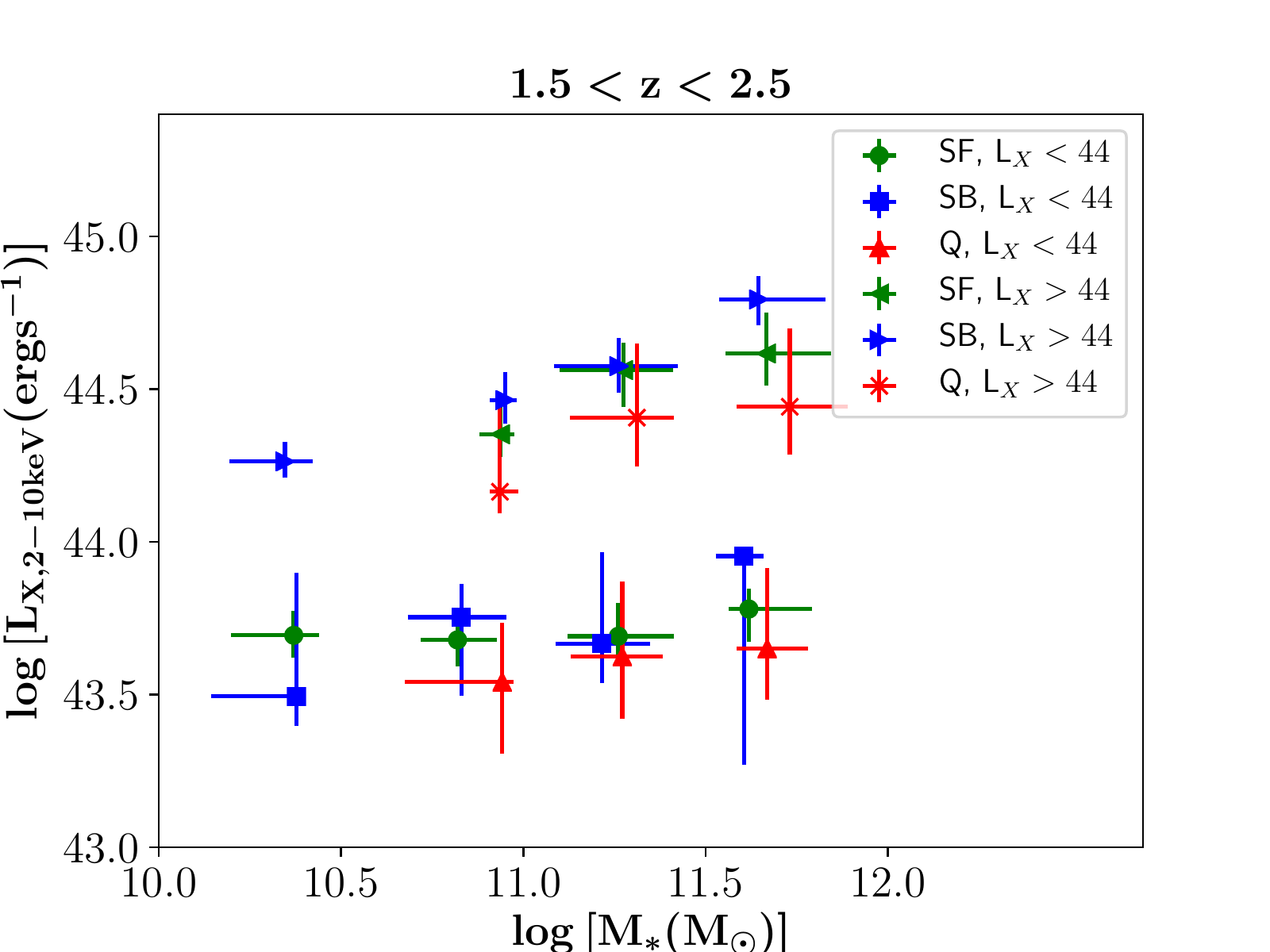}  
  \caption{L$_X$ as a function of M$_*$, for AGN with $\rm log\,[L_{X,2-10keV}(ergs^{-1})]>44$ and $\rm log\,[L_{X,2-10keV}(ergs^{-1})]<44$. The mild increase of L$_X$ with M$_*$ is mostly driven by the most luminous sources.}
  \label{fig_lx_mstar_lxsplit}
\end{figure}

Fig. \ref{fig_lx_mstar_lxsplit}, presents the L$_X$-M$_*$ relation for the two AGN populations. The (mild) increase of L$_X$ with M$_*$ seems to be mainly driven by the most luminous X-ray sources. 

\begin{figure}
\centering
  \includegraphics[width=0.8\columnwidth, height=6cm]{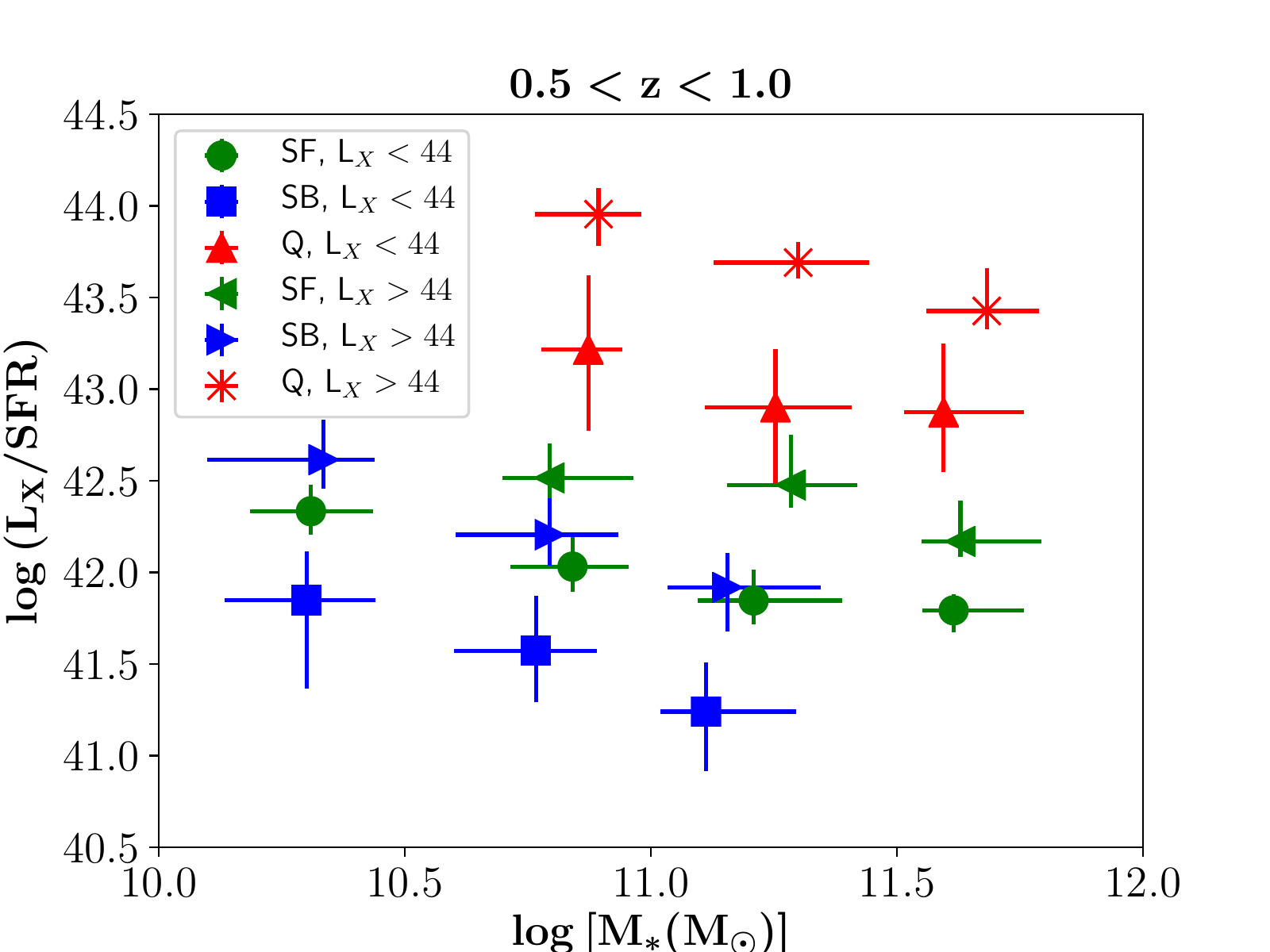} 
  \includegraphics[width=0.8\columnwidth, height=6cm]{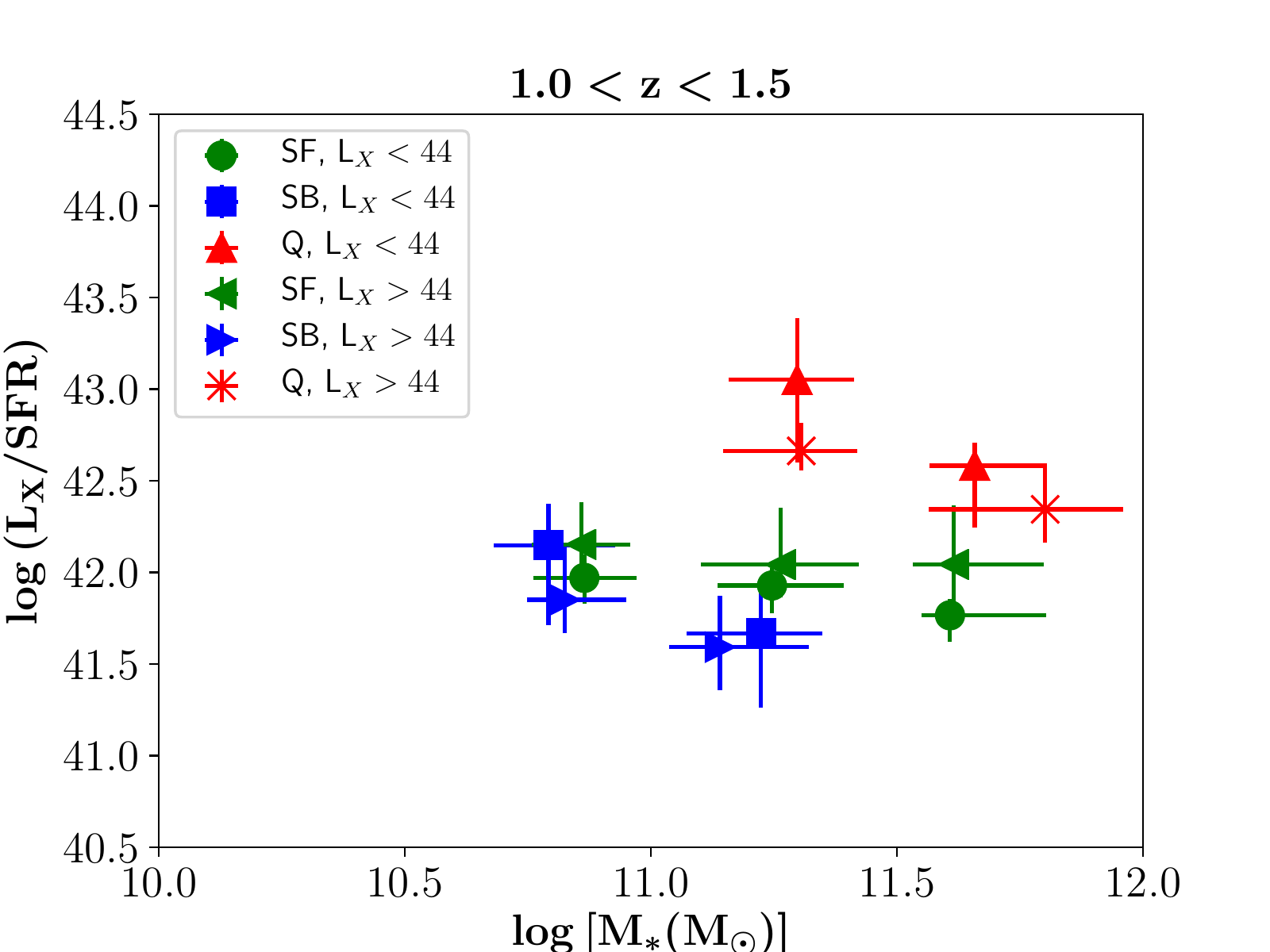} 
  \includegraphics[width=0.8\columnwidth, height=6cm]{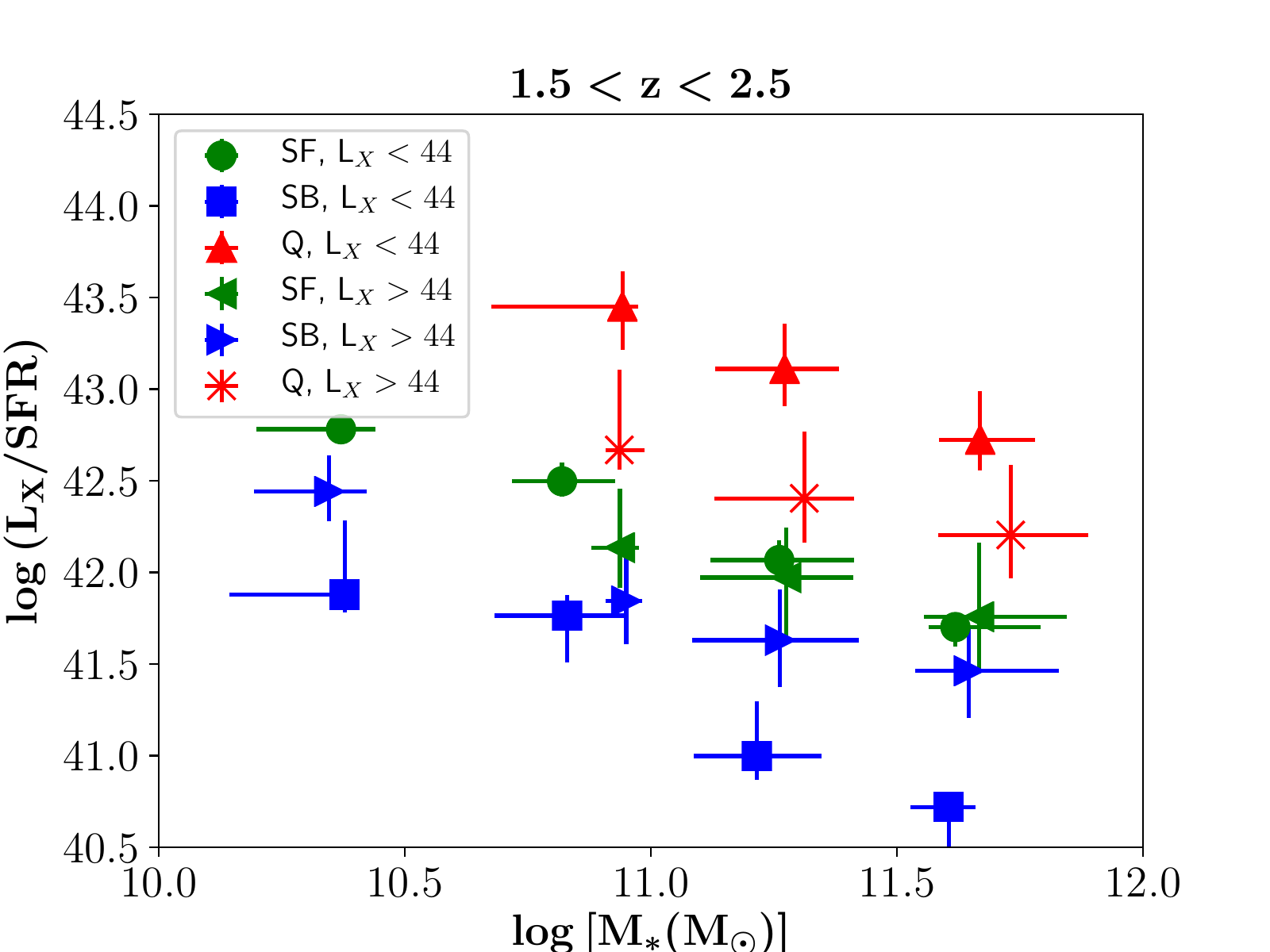}  
  \caption{The difference of the amplitude of L$_X$/SFR between luminous AGN ($\rm log\,[L_{X,2-10keV}(ergs^{-1})]>44$) and low-to-moderate luminosity AGN ($\rm log\,[L_{X,2-10keV}(ergs^{-1})]<44$). At $\rm 0.5<z<1.0$ luminous AGN have higher L$_X$/SFR values compared to their lower L$_X$ counterparts, for all AGN host galaxy classifications. At higher redshifts, luminous AGN hosted by Q systems present a  lower L$_X$/SFR compared to lower L$_X$ sources.}
  \label{fig_lxsfr_mstar_lxsplit}
\end{figure}

\begin{figure}
\centering
  \includegraphics[width=0.8\columnwidth, height=6cm]{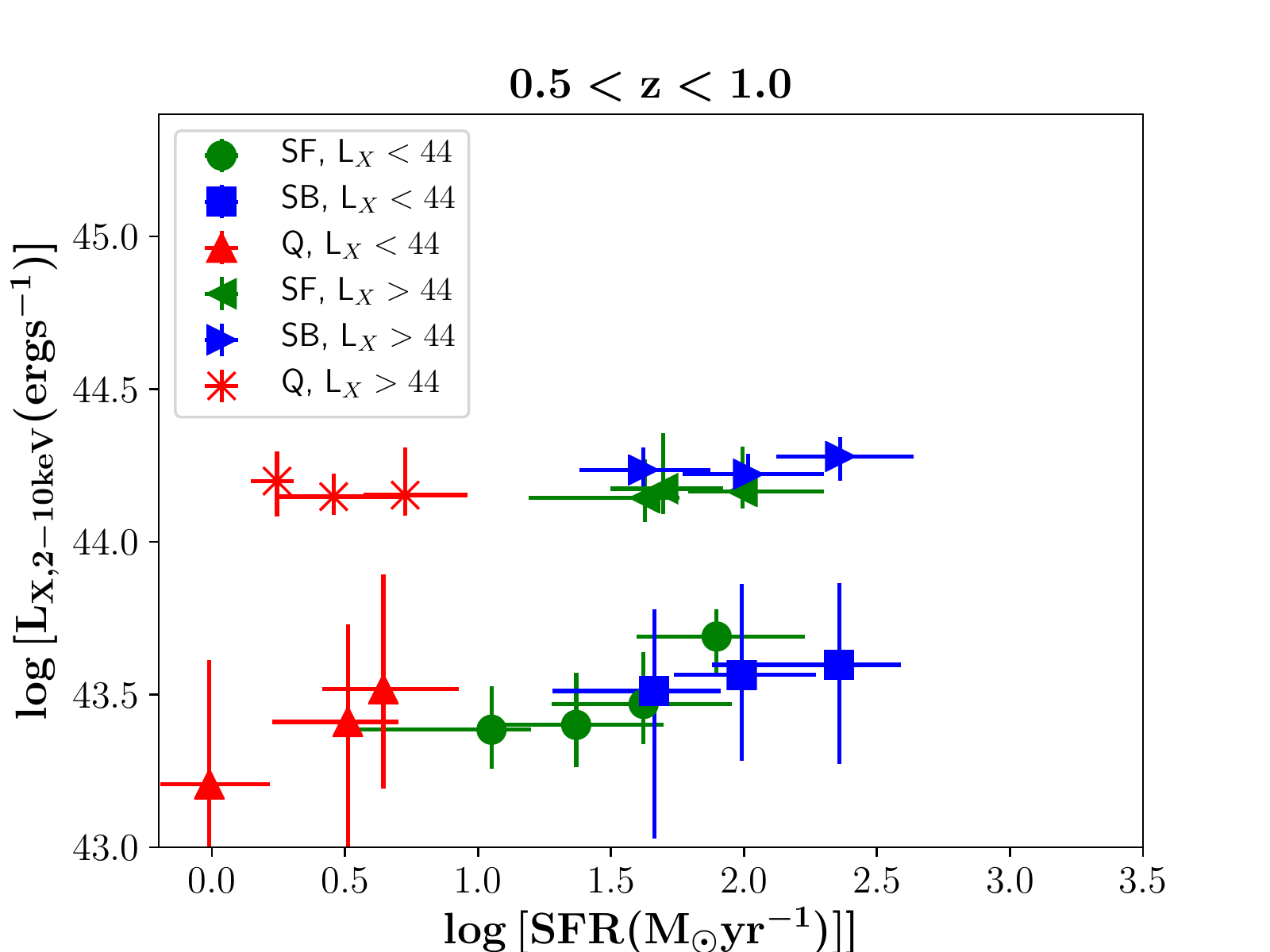} 
  \includegraphics[width=0.8\columnwidth, height=6cm]{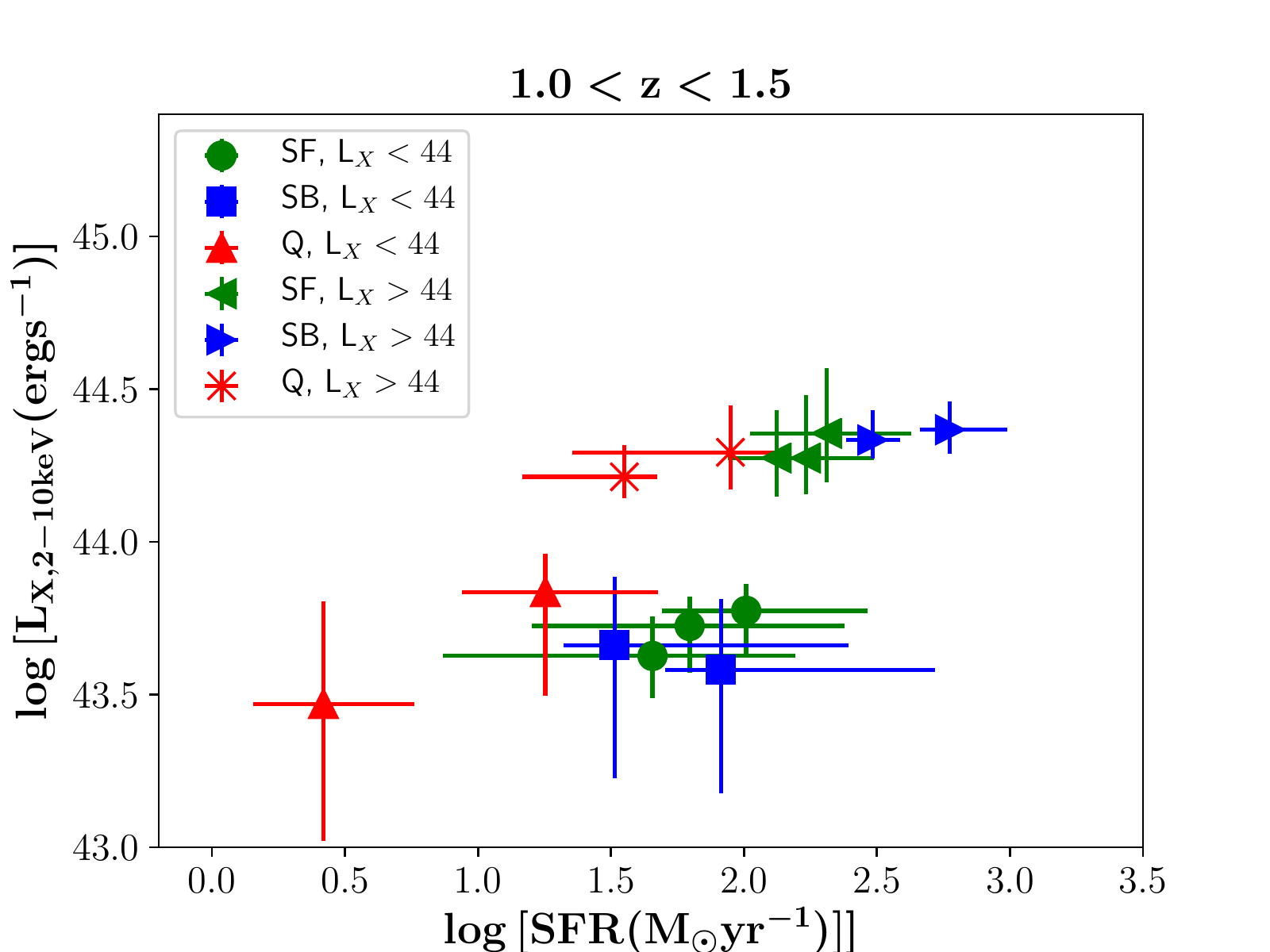} 
  \includegraphics[width=0.8\columnwidth, height=6cm]{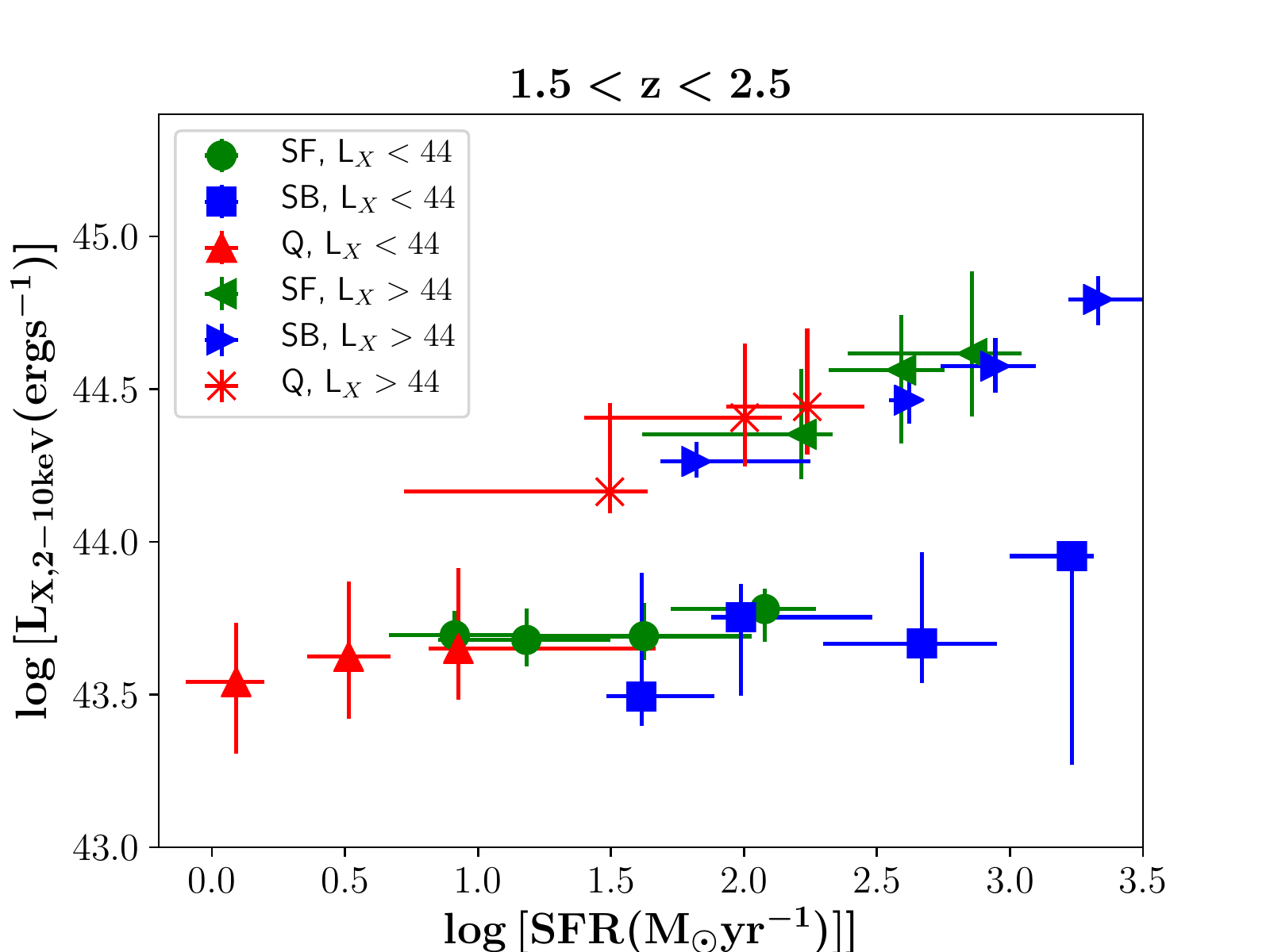}  
  \caption{L$_X$ as a function of SFR for SB, SF and Q AGN host galaxies, at three redshift intervals, for luminous and low-to-moderate L$_X$ AGN. At low redshifts (top panel), both AGN populations have similar SFR. However, at high redshifts (bottom panel) this is true only for the SB systems. } 
  \label{fig_lx_sfr_check_lxsplit}
\end{figure}

In Fig. \ref{fig_lxsfr_mstar_lxsplit}, we examine the ratio $\rm log(L_X/SFR)$ as a function of M$_*$, for luminous and low to moderate L$_X$ AGN. We find similar trends for both AGN populations, with luminous sources to have higher $\rm log(L_X/SFR)$ amplitude compared to less luminous X-ray sources, at low redshifts. However, at higher redshifts, luminous AGN hosted by Q galaxies present a lower $\rm log(L_X/SFR)$ amplitude compared to their lower L$_X$ counterparts. To investigate this behaviour further, in Fig. \ref{fig_lx_sfr_check_lxsplit}, we plot the L$_X$ vs. SFR for the two AGN populations. Based on these results, at low redshifts (top panel of Fig. \ref{fig_lx_sfr_check_lxsplit}), for all AGN host galaxy classifications, luminous and less luminous AGN have similar SFR. At higher redshifts (bottom panel of Fig. \ref{fig_lx_sfr_check_lxsplit}) this is true only for the SB galaxies, while Q systems that host less luminous AGN have significant lower SFR (by $\sim 1$\,dex) compared to their more luminous counterparts.

\end{document}